\documentclass[aps,epsf,preprint]{revtex4}
\usepackage{amssymb}

\def\1{{\bf 1}}
\def\[{\left[}
\def\]{\right]}
\def\be{\begin{eqnarray}}
\def\ee{\end{eqnarray}}
\def\bm{\begin{matrix}}
\def\em{\end{matrix}}
\def\nn{\nonumber}
\def\({\left(}
\def\){\right)}
\def\bk#1{\langle#1\rangle}
\def\eq#1{(\ref{#1})}

\def\a{\alpha}

\def\s{\sigma}
\def\o{\omega}

\def\f{\phi}

\def\G{{\cal G}}
\def\C{{\cal C}}
\def\l{\lambda}
\def\m{\mu}
\def\x{\times}

\def\bra#1{\langle #1|}

\def\d{\delta}

\def\labels#1{\label{#1}}
\def\edc{\end{document}}

\def\Rw{\Rightarrow}
\def\bn{\begin{enumerate}}

\def\en{\end{enumerate}}
\def\b{\beta}

\def\ol{\overline}
\def\rd{\sqrt{2}}
\def\rt{\sqrt{3}}

\def\rs{\sqrt{6}}
\def\diag{{\rm diag}}
\def\th{\theta}
\def\ba{\begin{array}}
\def\ea{\end{array}}
\def\bc{\begin{center}}
\def\ec{\end{center}}

\def\ol{\overline}

\def\edoc{\end{document}}

\begin{document}

\title{Horizontal Symmetries $\Delta(150)$ and $\Delta(600)$}
\author{C.S. Lam}
\address{Department of Physics, McGill University\\
 Montreal, Q.C., Canada H3A 2T8\\
and\\
Department of Physics and Astronomy, University of British Columbia,  Vancouver, BC, Canada V6T 1Z1 \\
Email: Lam@physics.mcgill.ca}
%\date{\today}

\begin{abstract}
Using group theory of mixing to examine all finite subgroups
of $SU(3)$ with an order less than 512, we found recently that
only the group $\Delta(150)$ can give rise to a correct
reactor angle $\th_{13}$ of neutrino mixing without any free parameter. It predicts $\sin^22\th_{13}=0.11$
and a sub-maximal atmospheric angle with $\sin^22\th_{23}=0.94$, in good agreement with experiment.
The solar angle $\th_{12}$, the CP phase $\d$, and the neutrino masses $m_i$ are left as free parameters.
In this article we provide more details of this case, discuss possible gain and loss by introducing 
right-handed symmetries, and/or valons to construct dynamical models. A simple model is discussed
where the solar angle agrees with experiment, and all its
mixing parameters can  be obtained
from the group $\Delta(600)$ by symmetry alone. 
The promotion of $\Delta(150)$ to $\Delta(600)$ is on the one
hand analogous to the promotion of $S_3$ to $S_4$ in the presence of tribimaximal mixing, and on the
other hand similar to the extension from $A_4$ to $S_4$ in that case.

\end{abstract}
%\pacs{}
\narrowtext
\maketitle

\section{Introduction}
Before the reactor angle $\th_{13}$ was successfully measured in the recent past, the neutrino mixing data were consistent with a zero $\th_{13}$ and a maximal atmospheric angle $\th_{23}$. These two are explainable
by the leptonic symmetry $S_3$. This symmetry group is generated by two unitary matrices, $F=\diag(1,\o,\o^2)$, where
$\o=\exp(2\pi i/3)$, and  $G$ with  $G^2=1$, $\det(G)=1$, which possesses
the eigenvector $(0,1,-1)^T$ with eigenvalue $+1$.
This symmetry predicts a zero reactor angle and a maximal
atomospheric angle, but  leaves the solar angle $\th_{12}$ undetermined.

The solar angle can also be explained by symmetry if we enlarge the group $S_3$ by incorporating 
another unitary generator $G'$ which commutes with $G$, satisfying ${G'}^2=1$, $\det(G')=1$,
and possessing an eigenvector $(1,1,1)^T$ with eigenvalue $+1$. The order of the
group $\G$ generated by $F,G,G'$ must then be an even multiple of 6, the order of $S_3$.

The simplest non-abelian group with order 12 is $A_4$, but  it contains only
even permutations so it does not contain $S_3$
as a subgroup. The simplest non-trivial group that contains $S_3$ is $S_4$, with order 24. Moreover,
it also contains the $G'$ above, so it predicts the correct solar angle. The 
resulting neutrino matrix has the well-known tribimaximal form.

Now that the reactor angle is no longer zero and there are indications that the atmospheric angle
may not be maximal, the question is whether we can still find a symmetry to explain the mixing
data. In a recent work, we searched all the finite subgroups of $SU(3)$ with an order less than 512, and discovered that the only group
that can explain the reactor and atomospheric angles by symmetry alone is $\Delta(150)$ \cite{LAM1}. 
It is generated by the same $F$ as $S_3$, but a different $G$.
It also leaves
the solar angle free. In order to have a hope to have the solar angle also explained by symmetry, 
we need to enlarge the group by incorporating another order-2 generator $G'$ which commutes with $G$.
As before, the extended group $\G$ must have an order which is an even multiple of 150, the order
of $\Delta(150)$. Again no non-trivial order-300 group contains a $\Delta(150)$ subgroup, but the
order-600 group $\Delta(600)$ does, and in fact that is the only order-600 group that does. The 
symmetry of $\Delta(600)$ also contains the same $G'$ before which 
gives rise to a solar angle that agrees with experiment.
In this way the pair ($\Delta(150), \Delta(600))$ is analogous to the previous pair $(S_3, S_4)$
before the reactor angle was measured. 

It is interesting that $\Delta(600)$ is the smallest of the three groups that yield a full $Z_2\x Z_2$
residual symmetry in the neutrino sector, among all the  finite $SU(3)$ subgroups of orders $<1536$ \cite{HLL}.
It also shows up in the method using von Dyck groups  \cite{HS, Hu}. 

In this article we  discuss  the groups $\Delta(150)$ and $\Delta(600)$ and their
predictions \cite{TRIUMF}. The mathematical
properties of these groups are summarized in Sec.~II. Given a symmetry group, there are three ways to apply
its symmetries to neutrino physics, each
with an additional assumption than the previous,
 but potentially (though not necessarily) also an additional gain. We will discuss these three  in turn.
The simplest is to apply symmetry only to the left-handed fermions, 
to obtain information on the effective left-handed
mass matrices. This is the method used in the recent survey \cite{LAM1} and shall be referred to as 
the `left-handed symmetry' method.
The next simplest is to apply symmetry to both the left-handed and the right-handed fermions,
to obtain information also on the Dirac and Majorana mass matrices. We shall refer to this method
as `both-handed symmetry'. Without imposing the additional right-handed symmetry there is
no way to get the Dirac or the Majorana mass matrices.
The
most involved but the most well-known method is to construct dynamical models. This requires the 
 introduction of  additional scalar fields called
valons to account for the spontaneous breaking of symmetry.
These three methods will be applied to $\Delta(150)$ and their relative merits will be discussed
in Secs.~III, IV, V respectively.

No matter which of these three methods is used, we can account for the correct reactor and 
atmospheric angles as long as $\Delta(150)$ is the symmetry. The other
neutrino mixing parameters, as well as the fermion masses, are hidden in the adjustable parameters
in the mass matrices. For left-handed symmetry, there are just enough parameters to fit all the experimental
quantities. For both-handed symmetry, the additional freedom in assigning right-handed symmetry
increases the number of parameters, but except for special cases such as
the $c=0$ example in the text, the additional parameters do not yield additional information on
the mixing parameters nor the masses. Nevertheless, when dynamical models are built, most of these
parameters in both-handed symmetry will acquire a physical meaning in terms of Yukawa coupling constants.
The groups $\Delta(6n^2)$ can also be incorporated into grand-unified-theory models \cite{IG}.

In principle, a fit to the experimental data can determine the parameters in left-handed symmetry, but 
in practice this is impossible because CP-phase and one of the neutrino masses are unknown. 
The best one can do then is to produce simple models, hopefully with simple choice of parameters,
that can account for all the known experimental quantities. Such a model will be given in the text,
first in Sec.~III, subsequently followed up and refined in Secs.~IV and V. 

In Sec.~VI, the left-handed symmetry
 will be applied to $\Delta(600)$, and in Sec.~VII, a summary of the results will be given.
There is also an appendix at the end to explain real representations for the group $\Delta(150)$.

\section{$\Delta(6n^2)$}

This series of finite subgroups of $SU(3)$ have been studied in various places \cite{DELTA}. 
In this article we will follow the information given in the mathematical software GAP \cite{GAP}.

$\Delta(6n^2)$ can be described by the semidirect products of a direct product of cyclic groups,
\be \Delta(6n^2)=(Z_n\x Z_n)\rtimes Z_3\rtimes Z_2.\labels{SD}\ee
The generators of the two cyclic groups $Z_n$ are $f_4$ and $f_3$ respectively; 
the generator of $Z_3$ is $f_2$, and the generator of $Z_2$ is $f_1$. The orders of $f_4, f_3, f_2, f_1$ are respectively
$n, n, 3, 2$.

The direct-product of the two $Z_n$'s tells us that
$f_4$ and $f_3$ commute. The semidirect with $Z_3$ implies that $f_2f_3f_2^{-1}$ and $f_2f_4f_2^{-1}$
are monomials of $f_3$ and $f_4$. The semidirect product with $Z_2$ implies that $f_1f_3f_1^{-1}$
and $f_1f_4f_1^{-1}$ are monomials of $f_2$, $f_3$ and $f_4$, while $f_1f_2f_1^{-1}$ is  a monomial of $f_2$. 

With these commutation relations, one can rearrange any monomial of the four $f_i$'s, namely any group element of $\Delta(6n^2)$, into the canonical form $f_4^{e_4}f_3^{e_3}f_2^{e_2}f_1^{e_1}$,
with $e_4, e_3$ ranging between 0 and $n-1$, $e_2$ between 0 and 2, and $e_1$ either 0 or 1. 
The total number of independent elements of the group can easily be counted from the canonical form to 
be $6n^2$.

In this article we are interested in the case $n=5$ and $n=10$. We will discuss the details of these two groups separately below.

\subsection{$\Delta(150)=(Z_5\x Z_5)\rtimes Z_3\rtimes Z_2$}
The nontrivial commutation relations
of the generators are
\be
 f_2f_4f_2^{-1}&=&f_4f_3,\quad       f_2f_3f_2^{-1}=f_4^2f_3^3,\nn\\
f_1f_4f_1^{-1}&=&f_4^4f_3^4,\quad       f_1f_3f_1^{-1}=f_3,\quad f_1f_2f_1^{-1}=f_2^2.\labels{f150}\ee

This group has 13 conjugacy classes, $\C_1$ to $\C_{13}$, and 13 irreducible representations, IR1 to IR13.
Its character table is given in Table 1, with various notation explained in Table 2 and below. 
The GAP notation $/A$ in Table 1 means the complex conjugate of $A$, and similarly  $/B, /F$, and 
$/G$. The notation $*C$ is obtained from $C$ by changing $\sqrt{5}$ to $-\sqrt{5}$, and similarly
 $*D$ and $*E$.

The designations of the IRs in Table 1 are those in the IrreducibleRepresentations command
of GAP. They differ
from those used in the command CharacterTable, which are denoted as X.n
in GAP, with n varying from 1 to 13.

The generators $f_1,f_2,f_3,f_4$ belong to classes
$\C_2, \C_3, \C_4, \C_5$ respectively. The only class with order 2 is $\C_2$. No two elements within this 
class mutually commute.

As usual, the dimension of an IR is its character for the identity class $\C_1$. 
The characters $\chi$ of the 3-dimensional representations IR4 to IR11 are complex, with the pairs 
(4,10), (5,11), (6,8), (7,9) being complex-conjugates of each other. The other representations,
IR1 to IR3 and IR12, IR13, have real characters, so they are either real or quaternionic representations. Their
Frobenius-Schur indices $\sum_{g\in\Delta(150)} \chi(g^2)/150$  are all $+1$, so all of  them
are real representations. The representations given by GAP for IR1 and IR2 are explicitly real, 
but not
those for IR3, IR12 and IR13. In Appendix A we show how to convert the GAP-representations into representations
that are explicitly real.

$$\ba{|c|ccccccccccccc|}\hline
\C&1&2&3&4&5&6&7&8&9&10&11&12&13\\ \hline
{\rm order}& 1 & 2 & 3 & 5 & 5 & 10 & 10 & 5 & 5 & 10 & 10 & 5 & 5 \\ \hline
|\C|&1 & 15 & 50 & 3 & 6 & 15 & 15 & 3 & 6 & 15 & 15 & 3 & 3 \\  \hline
{\rm IR1}& 1&  1&  1&  1&  1&   1&   1&  1&  1&   1&   1&  1&  1\\
{\rm IR2}&1& -1&  1&  1&  1&  -1&  -1&  1&  1&  -1&  -1&  1&  1\\
{\rm IR3}&2&  0& -1&  2&  2&   0&   0&  2&  2&   0&   0&  2&  2\\
{\rm IR4}& 3&  1&  0& /B& *D& -/G& -/F& /A&  D&  -F&  -G&  A&  B\\
{\rm IR5}&3& -1&  0& /B& *D&  /G&  /F& /A&  D&   F&   G&  A&  B\\
{\rm IR6}&3&  1&  0& /A&  D& -/F&  -G&  B& *D& -/G&  -F& /B&  A\\
{\rm IR7}& 3& -1&  0& /A&  D&  /F&   G&  B& *D&  /G&   F& /B&  A\\
{\rm IR8}&3&  1&  0&  A&  D&  -F& -/G& /B& *D&  -G& -/F&  B& /A\\
{\rm IR9}&3& -1&  0&  A&  D&   F&  /G& /B& *D&   G&  /F&  B& /A\\
{\rm IR10}&3&  1&  0&  B& *D&  -G&  -F&  A&  D& -/F& -/G& /A& /B\\
{\rm IR11}&3& -1&  0&  B& *D&   G&   F&  A&  D&  /F&  /G& /A& /B\\
{\rm IR12}& 6&  0&0&  C&  E&   0&0& *C& *E&  0&0& *C&  C\\
{\rm IR13}&6&  0&0&  *C&  *E&   0&0& C& E&  0&0& C&  *C\\
\hline\ea$$
\vskip.2cm
\bc Table 1. Character Table of $\Delta(150)$\ec
\vskip.5cm

$$\ba{|c|c|c|c|c|c|c|}\hline
A&B&C&D&E&F&G\\ \hline 
2\eta^3+\eta^4&\eta^2+2\eta^4&1-\sqrt{5}&(1-\sqrt{5})/2&(-3+\sqrt{5})/2& -\eta^4& -\eta^2\\
\hline\ea$$
\vskip.5cm
\bc Table 2. Abbreviations used in Table 1, where $\eta=e^{2\pi i/5}$\ec
\vskip.5cm

Using Table 1, we can explicitly work out the Clebsch-Gordan series. The result is given in Table 3.
Since IRi $\x$ IRj is the same as IRj $\x$ IRi, only half of the table is explicitly shown. For example, the products of IR12 and IR12 is
given by (with the prefix IR omitted)
\be
12\x12=1+2+3+3+6+7+8+9+12+12+13.\labels{1212}\ee
From Table 2, it can be checked that there are 36 states both on the lefthand side and the righthand side
of \eq{1212}.

{\footnotesize
$$\ba{| r| c| c| c| c| c| c| c| c| c| c| c| c|c|}\hline
&1&2&3&4&5&6&7&8&9&10&11&12&13\\ \hline
1&1&2&3&4&5&6&7&8&9&10&11&12&13\\
2&&1&3&5&4&7&6&9&8&11&10&12&13\\
3&&&1,2,3&4,5&4,5&6,7&6,7&8,9&8,9&10,11&10,11&12^2&13^2\\
4&&&&6,10,11&7,10,11&8,12&9,12&10,13&11,13&1,3,12&2,3,12&4,5,8,9,13&6,7,12,13\\
5&&&&&6,10,11&9,12&8,12&11,13&10,13&2,3,12&1,3,12&4,5,8,9,13&6,7,12,13\\
6&&&&&&8,9,10&8,9,11&1,3,13&2,3,13&4,13&5,13&10,11,12,13&4,5,6,7,12\\
7&&&&&&&8,9,10&2,3,13&1,3,13&5,13&4,13&10,11,12,13&4,5,6,7,12\\
8&&&&&&&&4,6,7&5,6,7&6,12&7,12&4,5,12,13&8,9,10,11,12\\
9&&&&&&&&&4,6,7&7,12&6,12&4,5,12,13&8,9,10,11,12\\
10&&&&&&&&&&4,5,8&4,5,9&6,7,10,11,13&8,9,12,13\\
11&&&&&&&&&&&4,5,8&6,7,10,11,13&8,9,12,13\\
12&&&&&&&&&&&&1,2,3^2,6,7,&4,5,6,7,8,9,\\
    &&&&&&&&&&&& 8,9,12^2,13                               &10,11,12,13\\
13&&&&&&&&&&&&&1,2,3^2,4,5\\
    &&&&&&&&&&&&&10,11,12,13^2\\
\hline\ea$$}
\vskip.2cm
\bc Table 3. Clebsch-Gordan series of $\Delta(150)$\ec

\subsection{$\Delta(600)=(Z_{10}\x Z_{10})\rtimes Z_3\rtimes Z_2$}
Since $Z_{10}=Z_5\x Z_2'$ (a prime is used to distinguish this $Z_2$ from the one in the semidirect
product), its generator $f_4$ is related to the $Z_5$ generator $c_4$ and the $Z_2'$ generator $d_4$ by
\be c_4={f_4}^2,\quad d_4={f_4}^5,\quad f_4={c_4}^3d_4.\labels{f4}\ee
Similarly, for the other $Z_{10}$ in the direct product,
\be c_3={f_3}^2,\quad d_3={f_3}^5,\quad f_3={c_3}^3d_3\labels{f3}.\ee
The nontrivial commutation relations between $f_1,f_2,c_3,c_4$ are the same as in \eq{f150}, namely,
\be
 f_2c_4f_2^{-1}&=&c_4c_3,\quad       f_2c_3f_2^{-1}=c_4^2c_3^3,\nn\\
f_1c_4f_1^{-1}&=&c_4^4c_3^4,\quad       f_1c_3f_1^{-1}=c_3,\quad f_1f_2f_1^{-1}=f_2^2.\labels{f600c}\ee
Together they generate the subgroup $\Delta(150)$ of $\Delta(600)$. 

The other nontrivial 
commutation relations of $\Delta(600)$ are
\be
 f_2d_4f_2^{-1}&=&d_3,\quad       f_2d_3f_2^{-1}=d_4d_3,\nn\\
f_1d_4f_1^{-1}&=&d_4d_3,\quad       f_1d_3f_1^{-1}=d_3.\labels{f600d}\ee
The group formed by $\{f_1,f_2,d_3\}$ (or $\{f_1,f_2,d_4\}$) is $S_4$, thus $\Delta(600)$ contains
$S_4$ as a subgroup. The subgroup formed by $\{f_2,d_3\}$ (or $\{f_2,d_4\}$) is $A_4$, 
and the subgroup formed by $\{f_1,d_3\}$ (or $\{f_1,d_4\}$) is the Klein four-group $Z_2\x Z_2$.
\footnote{While we follow the notations for the GAP generators of $\Delta(150)$,  a
different set is used for
$\Delta(600)$ to make it transparent that it
contains the subgroups $\Delta(150), A_4$, and $S_4$. If a bar on top denotes the GAP
generators for $\Delta(600)$, then the relations between the two sets  are
$ c_3=\bar f_5^3\bar f_6^4,\  c_4=\bar f_5^2\bar f_6^2;\quad  \bar f_5=c_3^4c_4^2,\
\bar f_6=c_3c_4;\quad 
         d_3=\bar f_3,\        d_4=\bar f_4$.
}

Substituting \eq{f4} and \eq{f3} into \eq{f600}, we get the nontrivial commutation relations with the $Z_{10}\x
Z_{10}$ generators to be
\be
 f_2f_4f_2^{-1}&=&f_4^6f_3,\quad       f_2f_3f_2^{-1}=f_4^7f_3^3,\nn\\
f_1f_4f_1^{-1}&=&f_4^9f_3^9,\quad       f_1f_3f_1^{-1}=f_3. \labels{f600}\ee

The group $\Delta(600)$ has 33 classes and 33 IRs, as listed in Tables 4 and 5. We have no need for
the character table nor the Clebsch-Gordan series in this article, so they will not be given here.

The classes containing order-2 elements are $\C_8$ and $\C_{24}$. Unlike $\Delta(150)$, it is possible
to find two mutually commuting order-2 elements in this group. The significance of this remark will be
explained in Sec.~VI.

$$\ba{|c|ccccccccccccccccc|}\hline
\C&1&2&3&4&5&6&7&8&9&10&11&12&13&14&15&16&17\\ \hline
{\rm order}&  1& 5& 5& 5& 5& 5& 5& 2& 10& 10& 10& 10& 10& 10& 10& 10& 10\\ 
|\C|&1& 6& 6& 3& 3& 3& 3& 3& 6& 6& 3& 6& 6& 3& 6& 6& 3\\   \hline\hline
\C&18&19&20&21&22&23&24&25&26&27&28&29&30&31&32&33&\\ \hline
{\rm order}&10& 10& 10& 10& 
  10& 3& 2& 10& 10& 10& 10& 4& 20& 20& 20& 20 &\\
|\C|&6& 6& 3& 6& 6& 200& 30& 
  30& 30& 30& 30& 30& 30& 30& 30& 30&\\
\hline\ea$$
\vskip.2cm
\bc Table 4. Conjugacy classes of $\Delta(600)$\ec
\vskip.5cm
$$\ba{|c|ccccccccccccccccc|}\hline
{\rm IR}&1&2&3&4&5&6&7&8&9&10&11&12&13&14&15&16&17\\ \hline
{\rm cc}&1&2&3&4&5&21&22&23&24&25&16&17&18&19&20&11&12\\
{\rm dim}&1 & 1 & 2 & 3 & 3 & 3 & 3 & 3 & 3 & 6 & 3 & 3 & 3 & 3 & 6 & 3 & 3 \\ \hline\hline
{\rm IR}&18&19&20&21&22&23&24&25&26&27&28&29&30&31&32&33&\\ \hline
{\rm cc}&13&14&15&6&7&8&9&10&26&27&29&28&30&33&32&31&\\
{\rm dim}&3 & 3 & 6 & 3 & 3 & 3 & 3 & 6 & 6 & 6 & 6 & 6 & 6 & 6 & 6 & 6&\\
 \hline\ea$$
\vskip.2cm
\bc Table 5. Irreducible representations of $\Delta(600)$\ec
\vskip.5cm

\section{Left-Handed Symmetry for $\Delta(150)$}
%In a recent publication \cite{LAM1}, finite subgroups of $SU(3)$ with an order less than 512 were searched for a mixing vector that agrees with the third column of the neutrino mixing matrix.  The only  group
%satisfying this requirement is $\G=\Delta(150)$, which gives  $\sin^22\th_{13}=0.11$
%and $\sin^22\th_{23}=0.94$, in agreement with experiment. 

In this section we first review the general theory of mixing based on a symmetry of effective left-handed
mass matrices (to be abbreviated as `left-handed symmetry')  \cite{Lam2}. We will also 
provide much more detailed information on the specific case of $\Delta(150)$.

Every neutrino mixing matrix $U$ has a natural $Z_2\x Z_2$ symmetry generated by
\be G_i=\pm u_iu_i^\dagger\mp u_ju_j^\dagger\pm u_ku_k^\dagger,\quad (i=1,2,3)\labels{Gi}\ee
where $u_1,u_2,u_3$ are the three columns of $U$, referred to 
as {\it mixing vectors} below, and $(i,j,k)$ is a permutation of (1,2,3). It is
easy to see that $G_i$ is unitary, $G_i^2=1$, and $G_iG_j=\pm G_k$. $G_i$ is a symmetry of the Majorana neutrino
mass matrix $\ol M_\nu=\ol M_\nu^T$ in the basis where the left-handed effective charged-lepton
matrix $\ol M_e:=M_e^\dagger M_e$ is diagonal. This is so because 
\be G_i^T\ol M_\nu G_i=\ol M_\nu\labels{GM}\ee
follows from  $\ol M_\nu=\sum_i m_iu_iu_i^T$, with $m_i$ being the masses of the active neutrinos
with Majorana phases.

This symmetry in the neutrino sector is accompanied by a $Z_n$ symmetry in the charged-lepton sector.
Since the  $\ol M_e$ is diagonal, every unitary diagonal $3\x 3$ matrix $F$ satisfying $F^\dagger\ol M_eF=\ol M_e$  is  a symmetry. If $F^n=1$, then $F$ generates the said $Z_n$ symmetry.
$F$ cannot be identical to any  $G_i$, otherwise the mixing column $u_i$ would have
two zeros, contrary to experiments.

$\ol M_\nu$ and $\ol M_e$ are the (effective) left-handed mass matrices mentioned earlier.

So far everything is general. The group theory of mixing is based on the  assumption that
$F$ and at least one $G_i$ (hereafter referred to simply as $G$) originate from  an unbroken nonabelian
symmetry $\G$, and remain a symmetry after the spontaneous symmetry breaking, although the other members of $\G$ may not.
From now on
we shall assume $\G$ to be a finite group.

The minimal symmetry group $\G$ consistent with this assumption is generated by $F$ and $G$.
Conversely, given any $\G$, we can construct  mixing vectors $v$ (one of the $u_i$ above)
 in the following way.

Collect from $\G$ all possible pair of elements $(F,G)$ belonging to all possible 3-dimensional
irreducible representations in which $F$ and $G$ are unitary. 
In order to ensure $\ol M_e$ to be diagonal when $F$ is, we require
the three eigenvalues of $F$ to be distinct, hence its order $n$ must be at least 3. $G$ should be of order
2 because $G^2=1$. 
Let $v=(v_1,v_2,v_3)^T$ be the normalized eigenvector of $G$ with eigenvalue $-{\rm Tr}(G)\  (=\pm 1)$ in
the $F$-diagonal basis, then $v$ is a possible mixing vector and it is uniquely determined from $G$. In particular, if 
it is the mixing vector for the third column of the
neutrino mixing matrix $U$, then  $|v_1|=|\sin\th_{13}|,\  |v_2|=|\cos\th_{13}\sin\th_{23}|$, and $|v_3|=
|\cos\th_{13}\cos\th_{23}|$. Since the CP and Majorana phases are unknown, many distinct $v$'s
can give rise to the same $|v_i|$ which is all that can be measured. Moreover, by rearranging the
entries of the diagonal $F$, the components of $v_i$ are similarly rearranged. We shall refer to 
these distinct $v$'s that give rise to the same $|v_i|$ with some rearrangement as {\it equivalent mixing vectors}.
For experimental comparisons they need not be distinguished.

In addition to the $G$ whose mixing vector $v$  fits the third column of $U$, if we can find another $G'$ commuting with $G$ whose mixing vector $v'$ fits the first or the second column, then the unitarity of $U$
ensures the remaining column to agree with experiment. In this way we
could get a purely symmetric explanation
of all the mixing parameters, without invoking any free parameter. In other words, we have then 
a full $Z_2\x Z_2$ rather than a single $Z_2$ symmetry in the neutrino sector. A necessary condition
for this full symmetry to happen is to have two mutually commuting order-2 elements $G, G'$
in the group. As remarked in the last section, this can never happen in $\Delta(150)$, but it does
happen in $\Delta(600)$.

There are clearly many $(F,G)$ pairs for each group. The choice in $G$ is somewhat
 limited because it must belong to a conjugacy class of order 2, but $F$ can come from essentially everything else except the identity.
Fortunately, it is sufficient to take only one $F$ per conjugacy class of order $\ge 3$, provided
we choose $G$ to be all possible elements of order 2. This is so because the pair $(F,G)$ and the 
pair $(gFg^{-1}, gGg^{-1})$, for any $g\in\G$, give equivalent mixing vectors $v$, so other $F$'s
in the same conjugacy cannot yield an inequivalent mixing vector.

Let us now specialize to $\G=\Delta(150)$. As shown in Tables 1, it has 13 classes $\C_i$ and 
13 irreducible representations (IR). $G$ must be taken from $\C_2$, with $|\C_2|=15$,
but $F$ can be taken from
$\C_3$ to $\C_{13}$, hence there are 11 possible $F$'s. There are eight 
3-dimensional irreducible representations (IR4 to IR11), so altogether there are $15\x 11\x 8=1320$
distinct $(F,G)$ pairs. Most of these pairs do not yield an acceptable mixing vector for the third column,
but many pairs do, all giving equivalent mixing vectors $|v|=(.170, .607, .777)^T$,
corresponding to $\sin^22\th_{13}=0.11$ and $\sin^22\th_{23}=0.94$. These pairs occur in
IR$n$, with $F$ in $\C_i$, and $G$ being the $k$th element of $\C_2$. Table 6 consists all these
possible triplets $(n,i,k)$.

$$\ba{|ccc|ccc|ccc|ccc|}\hline
n&i&k&n&i&k&n&i&k&n&i&k\\ \hline
4& 3&  5&  5& 3&  5 & 6& 3&  3&  7& 3&  3 \\
 8& 3&  9&  9& 3&  9 & 10& 3&  6&  11& 3&  6 \\
4& 3&  6&  5& 3&  6 & 6& 3&  9&  7& 3&  9 \\
 8& 3&  3&  9& 3&  3 & 10& 3&  5&  11& 3&  5 \\
4& 3&  10&  5& 3&  10 & 6& 3&  14&  7& 3&  14 \\
 8& 3&  8&  9& 3&  8 & 10& 3&  12&  11& 3&  12 \\
4& 3&  11&  5& 3&  11 & 6& 3&  7&  7& 3&  7 \\
 8& 3&  15&  9& 3&  15 & 10& 3&  13&  11& 3&  13 \\
4& 3&  12&  5& 3&  12 & 6& 3&  8&  7& 3&  8 \\
 8& 3&  14&  9& 3&  14& 10& 3&  10&  11& 3&  10 \\
4& 3&  13&  5& 3&  13 & 6& 3&  15&  7& 3&  15 \\
 8& 3&  7&  9& 3&  7 & 10& 3&  11&  11& 3&  11\\
\hline\ea$$
\vskip.5cm
\bc Table 6. Origin of equivalent mixing vectors
\ec

We see from Table 6 that the successful $F$ comes from $\C_3$, whose GAP-basis expression turns out to be
{\small\be F=\pmatrix{ 0& 0& 1 \cr 1& 0& 0 \cr 0& 1& 0 \cr}\labels{F}\ee}
for all the IRs in Table 6. It can be diagonalized by $V$, yielding $V^\dagger FV=\diag(\o^2,\o,1)$,
with $\o:=\exp(2\pi i/3)$, where
{\small\be V={1\over\rt}\pmatrix{  \o & \o^2 & 1\cr \o^2 & \o & 1 \cr 1& 1& 1 \cr}.\labels{V}\ee}
If we permute the columns of $V$, then the entries of the diagonal form of $F$ are similarly
permuted.

The mixing vector $\tilde v$ {\it in the} GAP {\it basis} is the eigenvector of $G$ in that basis,
with eigenvalue $-{\rm Tr}(G)$. The mixing vector $v$ in the 
$F$-diagonal basis which yields a possible column of $U$ is related to it by $v=V^\dagger\tilde v$.

It turns out that there are only six different $G$ and $\tilde v$ pairs, to be labelled by $a,b,c,d,e,f$ below.
$a$ comes from the first two rows of Table 6, $b$ the next rows, so on, and $f$ comes from the last
two rows. They are
{\small\be
G_a&=&\pm\pmatrix{0&\eta^3&0\cr \eta^2&0&0\cr 0&0&1\cr}, \
G_b=\pm\pmatrix{0&\eta^2&0\cr \eta^3&0&0\cr 0&0&1},\
G_c=\pm\pmatrix{0&0&\eta^2\cr 0&1&0\cr \eta^3&0&0\cr},\nn\\ \nn\\
G_d&=&\pm\pmatrix{1&0&0\cr 0&0&\eta^3\cr 0&\eta^2&0\cr},\
G_e=\pm \pmatrix{0&0&\eta^3\cr 0&1&0\cr \eta^2&0&0\cr},\
G_f=\pm\pmatrix{1&0&0\cr 0&0&\eta^2\cr 0&\eta^3&0}, \labels{GGG}\ee}
where $\eta:=\exp(2\pi i/5)$. There are four big columns in Table 6. The $G$'s coming from
columns 1 and 3 have $\det(G)=-1$, and those coming from columns 2 and 4 have $\det(G)=+1$.

The corresponding unnormalized $\tilde v$  are 
{\small\be
\tilde v_a&=&\pmatrix{-\eta^3\cr 1\cr 0},\ \tilde v_b=\pmatrix{-\eta^2\cr 1\cr 0},\
\tilde v_c=\pmatrix{-\eta^2\cr 0\cr1},\nn\\ \nn\\  \tilde v_d&=&\pmatrix{0\cr-\eta^3\cr 1},\
\tilde v_e=\pmatrix{-\eta^3\cr 0\cr 1}, \tilde v_f=\pmatrix{0\cr -\eta^2\cr 1},\labels{vtilde}\ee}
The corresponding normalized mixing vectors $v=V^\dagger\tilde v$ are
{\small\be v_a&=&{1\over\rs}\pmatrix{\o(1-\o\eta^3)\cr\o^2(1-\o^2\eta^3)\cr1-\eta^3\cr},\
 v_b={1\over\rs}\pmatrix{\o(1-\o\eta^2)\cr\o^2(1-\o^2\eta^2)\cr1-\eta^2\cr},\
 v_c={1\over\rs}\pmatrix{1-\o^2\eta^2\cr 1-\o\eta^2\cr 1-\eta^2\cr },\nn\\ \nn\\
 v_d&=&{1\over\rs}\pmatrix{ 1-\o\eta^3\cr 1-\o^2\eta^3\cr 1-\eta^3\cr},\
 v_e={1\over\rs}\pmatrix{ 1-\o^2\eta^3\cr\ 1-\o\eta^3\cr 1-\eta^3\cr},\
 v_f={1\over\rs}\pmatrix{ 1-\o\eta^2\cr 1-\o^2\eta^2\cr 1-\eta^2\cr}.
\labels{v}\ee}
Since complex conjugation interchanges $\o$ with $\o^2$, and $\eta^2$ with $\eta^3$, and since
$|v|=|v^*|$, we conclude that 
$|v_a|=|v_c|=|v_d|$ and $|v_b|=|v_e|=|v_f|$. Moreover, $|v_b|$ is obtained from $|v_a|$
by interchanging the last two rows.
Hence
all six cases give rise to the same equivalent mixing vector whose normalized magnitude
is $|v|=(.170, .607, .777)^T$.

The mass matrix $\ol M_e=\ol M_e^\dagger$ can be obtained from the symmetry condition $F^\dagger
\ol M_e F$ to be 
{\small\be
\overline M_e=\pmatrix{
\alpha&\beta&\beta^*\cr \beta^*&\alpha&\beta\cr \beta&\beta^*&\alpha\cr} ,\labels{olme}\ee}
where $\a$ is a real parameter and $\beta$ a complex parameter. The masses $m_e^2, m_\mu^2$
and $m_\tau^2$ can be obtained from $\alpha, \beta_R:=\Re(\beta)$ and $\b_I:=\Im(\beta)$
to be $\a-\b_R+\rt \b_I, \a-\b_R-\rt\b_I, \a+2\b_R$.

Similarly, the neutrino mass $\ol M_\nu=\ol M_\nu^T$ can be obtained from the symmetry condition
$G^T\ol M_\nu G=\ol M_\nu$. Corresponding to the six $G_a$, we get respectively
{\small\be \ol M_{\nu a}&=&\pmatrix{
a&b&c\cr b&a\eta&c\eta^3\cr c&c\eta^3&f\cr},\
\ol M_{\nu b}=\pmatrix{
a&b&c\cr b&a\eta^4&c\eta^2\cr c&c\eta^2&f\cr},\
\ol M_{\nu c}=\pmatrix{
a&b&c\cr b&f&b\eta^2\cr c&b\eta^2&a\eta^4\cr},\nn\\ \nn\\
\ol M_{\nu d}&=&\pmatrix{
a&b&b\eta^3\cr b&f\eta^4&c\cr b\eta^3&c&f\cr},\
\ol M_{\nu e}=\pmatrix{
a&b&c\cr b&f&b\eta^3\cr c&b\eta^3&a\eta\cr},\
\ol M_{\nu f}=\pmatrix{
a&b&b\eta^2\cr b&f\eta&c\cr b\eta^2&c&f\cr},\labels{olmnu}\ee}
where $a,b,c,f$ are complex parameters which can be used to fit the solar angle, the CP phase,
the three neutrino masses and the three Majorana phases (one of them is an overall phase).

\subsection{A Simple Model}
From now on we will study case $a$ in more detail, and will write $\ol M_{\nu a}$ simply as $\ol M_\nu$.
As has already been remarked, the four complex parameters $a, b, c, f$ are just enough to fit the
neutrino masses and Majorana phases, the solar angle and the CP phase. In this subsection, we consider
a simple model with $c=0$, which turns out to give the right solar angle as well. The remaining
parameters $a, b, f$ can be used to determine the neutrino masses $m_i$ and the Majorana phases. 

With $c=0$, the mass matrix becomes
\be (\ol M_\nu)_0=\pmatrix{
a&b&0\cr b&a\eta&0\cr 0&0&f\cr}.\labels{olmnu0}\ee
It can be diagonalized by the matrix 
{\small\be W_0={1\over\rd}\pmatrix{\eta^3&0&-\eta^3\cr 1&0&1\cr 0&\rd&0},\labels{w0}\ee}
resulting in
{\small\be
W_0^T(\ol M_\nu)_0W_0=\pmatrix{a\eta-b\eta^3&0&0\cr  0&f&0\cr 0&0&a\eta+b\eta^3\cr}.\labels{modeldiag}\ee}
The neutrino mixing matrix in this case is
\be U_0=V^\dagger W_0={1\over\sqrt{6}}\pmatrix{\o+\o^2\eta^3&\rd&\o-\o^2\eta^3\cr
\o^2+\o\eta^3&\rd&\o^2-\o\eta^3\cr 1+\eta^3&\rd&1-\eta^3\cr},\labels{U0}\ee
where $V$ is the matrix in \eq{V} used to diagonalize $F$.

The third column of $U_0$ is just $v_a$ of \eq{v}, which gives the desired reactor and atmospheric angles.
The second column gives trimaximal mixing, yielding a solar angle $\th_{12}$ satisfying
\be \sin^22\theta_{12}=0.90,\labels{solar}\ee
consistent with the PDG value \cite{PDG}  $\sin^22\th_{12} = 0.95\pm 0.10\pm 0.01$.
The first column is determined from the other two by unitarity of $U_0$.

The CP phase can also be read off from the (21), (22), (31), or (32) element of 
$|(U_0)_{ij}|$. The result is $\delta_{CP}=0$. 

The neutrino masses are the absolute values of the diagonal elements in \eq{modeldiag}:
\be
m_1=|a\eta-b\eta^3|,\quad m_2=|f|,\quad m_3=|a\eta+b\eta^3|.\labels{numass}\ee
%Thus the atmospheric gap is given by
%\be \Delta m^2_{atm}=|m_3^2-m_1^2|=4 \Re(ab^*\eta^3),\labels{atmgap}\ee
%and the mid-point of the gap is located at
%\be (m_3^2+m_1^2)/2=|a|^2+|b|^2.\labels{midatm}\ee
%Since we do not know the mass values, the only other qualitative restriction is that the solar gap must be much
%less than the atmospheric gap, resulting in the inequality
%\be  \left|\ |f|^2-|a-b\eta^2|\ \right|\ll 4 \Re(ab^*\eta^3).\ee

There are enough parameters to accommodate the measured mass gaps, either with a normal hierarchy or an inverted hierarchy.
In fact, it can fit any Majorana phase as well. For example, if it should turn out that the Majorana phases are all zero, then
we would take $a\eta\mp b\eta^3$ and $f$ to be positive. In a normal hierarchy with $m_1=0$, then 
$m_2^2=f^2$ is the solar gap and $m_3^2=4|b|^2$ is the atmospheric gap. We can do so similarly for an inverted hierarchy.

The charged-lepton mass squares are given below (16) to be  $m_e^2=\a-\b_R+\rt\b_I,\ m_\m^2=\a-\b_R-\rt\b_I$,
and $m_\tau^2=\a+2\b_R$. Thus $m_\mu^2-m_e^2=-2\rt\b_I$ tells us that $\b_I$ is negative, $\a=(m_e^2+m_\mu^2+m_\tau^2)/3$ is positive, and $2\b_R=m_\tau^2-\a$ is also positive.

\section{Both-Handed Symmetry for $\Delta(150)$}
The left-handed mass matrices $\ol M_e=M_e^\dagger M_e$
and $\ol M_\nu=M_\nu^TM_N^{-1}M_\nu$ given in \eq{olme} and \eq{olmnu}
are the most general that yield the correct reactor and atmospheric angles from $\Delta(150)$.
They have just enough free
parameters to fit the unknown quantities: charged-lepton and neutrino masses, the solar angle, as well as
the CP and the Majorana phases. Left-handed symmetry alone
cannot determine these parameters, nor can they tell us what the Dirac mass matrices
$M_e, M_\nu$ and the Majorana mass matrix $M_N$ are. For the latter we need to impose additional
assumptions regarding how the right-handed leptons transform.

Suppose the right-handed charged leptons transform according
 to a (irreducible or reducible) representation $C'$, and
the right-handed Majorana neutrinos $N$ according to a representation $C$. Let
 $F_{C'}$, $G_C$ be the unitary matrices for the residual
symmetries in these representations. With these symmetries,
 the Dirac and Majorana mass matrices should obey the constraint
\be
M_e&=&F_{C'}^\dagger M_e F,\quad M_\nu=G^T_C M_\nu G,\quad
M_N=G_C^TM_NG_C,\labels{diracmaj}\ee
where $F$ is given in \eq{F} and $G$ is one of the six equivalent forms in \eq{olmnu}. 
In what follows
we shall confine ourselves to case $a$, namely,
{\small\be F=\pmatrix{ 0& 0& 1 \cr 1& 0& 0 \cr 0& 1& 0 \cr},\quad G=-\pmatrix{0&\eta^3&0\cr \eta^2&0&0\cr 0&0&1\cr}, \quad \tilde v=\pmatrix{-\eta^3\cr 1\cr 0},\labels{FGv}\ee}
and
{\small\be 
\overline M_e=\pmatrix{
\alpha&\beta&\beta^*\cr \beta^*&\alpha&\beta\cr \beta&\beta^*&\alpha\cr} ,\quad
\ol M_{\nu }=\pmatrix{
a&b&c\cr b&a\eta&c\eta^3\cr c&c\eta^3&f\cr}.
\labels{olmenu}\ee}
These matrices  are taken from IR5, with $F$ coming from class $\C_3$ and $G$ from $\C_2$. More
specifically, in terms of the generators $f_i$ discussed in Sec.~II, 
\be
F=f_2,\quad  G=f_1f_3^2f_4^4.\labels{FGf}\ee

In order for
the left-handed matrices $\ol M_e=M_e^\dagger M_e$
and $\ol M_\nu=M_\nu^TM_N^{-1}M_\nu$ to be given by \eq{olmenu},
$G_C$ must be real as well as unitary, a requirement which can also be seen from the Majorana nature of $N$.
 If $N$ is described by a Majorana field $\psi$, then
$\psi_c={\cal C}\ol\psi=\psi$, where ${\cal C}$ is the charge conjugation operator. 
To preserve the Majorana nature under an internal symmetry transformation
$\psi\to R\psi$, it is necessary to have $R=R^*$ because $\psi_c\to R^*\psi_c$.
As a result, $G=G^T$ because $G$ is unitary and of order 2, and $M_N^{-1}$ transforms like $M_N$:
\be M_N^{-1}=G^TM_N^{-1}G.\labels{olmenu2}\ee

With the introduction of right-handed symmetry, we expected  more parameters to appear in
$M_e, M_\nu$, and $M_N$ than in $\ol M_e$ and $\ol M_\nu$. Nevertheless, when we use the former three
to calculate the latter two, we must still get back to \eq{olmenu} and the parametrizations there. On the
one hand, since there are more parameters in the former than in 
the latter, we expect the resulting parameters in
the latter to remain independent. On the other hand, the composite nature of the latter 
may produce occasional surprises arising from special composite features. A case in point which
we will discuss later occurs in 
the model in Sec.~IIIA, where $c=0$.
 There are four different right-handed assignments giving the same
$\ol M_\nu$, and in three of the four cases, $m_3$ is forced to be zero as well.

If future experiment should reveal that $m_3\not=0$, then this is a strong support for the symmetry 
assignment of $N$ to be the fourth. This 
conclusion is what we gain by imposing the right-handed symmetry. 

We will assume  fermions of different generations to be distinguished by different family quantum numbers,
hence if $C'$ and $C$ are reducible, the irreducible representations they contain  must not repeat
themselves. Also, included in the latitude of choice of $C$ is the number $\rho$ of $N$'s, which is
unknown. We reject $\rho=1$ to retain the possibility of leptogenesis, and will confine
ourselves here to $\rho=2$ and 3.

\subsection{Charged-Lepton Sector}
There are three right-handed charged leptons, so the dimension of $C'$ must be 3. If it is irreducible,
from Table 1 we see that it must belong to one of IR4 to IR11. If it is reducible, it could either be IR3+IR1
or IR3+IR2.

It turns out that $F_{C'}=F$ if $C'$ is one of IR4 to IR11. Even so, the solution of \eq{diracmaj} for $M_e$
is not the same as  solution  $\ol M_e$ in \eq{olmenu} because $\ol M_e$ has to be hermitian but
$M_e$ may not be. The general solution of \eq{diracmaj} for $M_e$ turns out to be
{\small\be
M_e=\pmatrix{ \zeta&\sigma&\tau\cr \tau&\zeta&\sigma\cr \sigma&\tau&\zeta\cr}.\labels{mea}\ee}
If it is hermitian, then $\tau=\sigma^*$ and we get back to the form of $\ol M_e$ in \eq{olmenu}.
In any case, we can compute $\ol M_e=M_e^\dagger M_e$ from \eq{mea} and obtain \eq{olmenu} with
$\alpha=|\zeta|^2+|\sigma|^2+|\tau|^2$ and $\beta=\zeta^*\sigma+\sigma^*\tau+\tau^*\zeta$.

Both $\ol M_e$ and $M_e$ give the same physics, so how come the former is described by three 
real parameters, $\a, \b_R, \b_I$, but the latter is expressed in terms of three complex parameters
$\zeta, \s, \tau$? This puzzle can be solved by computing $V^\dagger M_e V$ to diagonalize $M_e$. 
One  finds  its three eigenvalues to be complex. Their magnitudes are the masses and their
phases are unphysical and can be absorbed into the right-handed charged-lepton fields.

Next, suppose $C'$ is IR3+IR1 or IR3+IR2. Then using the real representations given in \eq{bf3}, we obtain
{\small\be F_{C'}={1\over 2}\pmatrix{  -1&-\rt&0\cr \rt&-1&0\cr 0&0&2\cr}\labels{fr}\ee}
for both cases. The symmetry constraint \eq{diracmaj} then yield
{\small\be M_e=\pmatrix{\zeta&\s&-\zeta-\s\cr -(2\s+\zeta)/\rt& (\s+2\zeta)/\rt&(\s-\zeta)\rt\cr\tau&\tau&\tau\cr}.\labels{meb}\ee}
 If we compute $\ol M_e=M_e^\dagger M_e$ from \eq{meb},  we get back to \eq{olmenu} with
$\alpha=2(|\zeta|^2+|\sigma|^2+|\zeta+\sigma|)^2/3+|\tau|^2$ and $\beta=2(-|\zeta+\s|^2-\s^*\zeta)/3
+|\tau|^2$.

\subsection{Neutrino Sector}
 As mentioned before,
the representation of a Majorana neutrino must be real. Since IR5 is complex, it is not possible
to realize $\ol M_\nu$ in a type-II seesaw mechanism. There is no problem to realize it with a
type-I seesaw, as implicitly assumed before, as long as the representation
$C$ of $N$ is real. 
The only real representations of $\Delta(150)$ (see Table 1 and Appendix A) are IR1, IR2, IR3, IR12 and IR13.
The dimensions of IR12 and IR13 are six, so we can ignore them if $\rho<6$. This leaves only two
possibilities each for $\rho=2$ and 3. For $\rho=2$, $C=$IR3 or IR1+IR2. For $\rho=3$, $C$ is either
IR3+IR1 or IR3+IR2. In fact, if $\rho<6$, the only other possibility is $\rho=4$ with $C$=IR3+IR2+IR1,
but that contains many parameters and sheds no light on the existing physics so we will consider 
it further.

We consider these four cases separately below.

\subsubsection{\underline{$\rho=2,\ C=$ IR3}}
From \eq{FGf}, Table 1 and \eq{bf3}, we get
{\small\be
G_C=\ol f_1(3)\ol f_3(3)^2\ol f_4(3)^4=\pmatrix{0&1\cr 1&0\cr}
\labels{G3}\ee}
The solution of \eq{diracmaj} is 
{\small\be M_N^{-1}=\pmatrix{P&Q\cr Q&P\cr},\quad M_\nu:=\pmatrix{x&y&z\cr -y\eta^2&-x\eta^3&-z\cr}.
\labels{case1}\\ \ee}
Computing from these $\ol M_\nu=M_\nu^TM_N^{-1}M_\nu$, we obtain \eq{olmenu} with
\be
a&=&(x^2+y^2\eta^4)P-2xy\eta^2Q,\quad b=2xyP-(x^2+y^2\eta)\eta^2Q,\nn\\
c&=&z(x+y\eta^2)(P-Q),\quad f=2z^2(P-Q),\quad\Rw\nn\\
a+b\eta^2&=&(x+y\eta^2)^2(P-Q).\labels{abcf3}\ee

\subsubsection{\underline{$\rho=2,\ C=$ IR1+IR2}}
From Table 1, we get
{\small\be G_C=\pmatrix{1&0\cr 0&-1\cr}.\labels{G12}\ee
The solution of \eq{diracmaj} is 
{\small\be M_N^{-1}=\pmatrix{P&0\cr 0&Q\cr},\quad M_\nu:=\pmatrix{x&-x\eta^3&0\cr y&y\eta^3&z\cr}.
\labels{case2}\\ \ee}
Computing from these $\ol M_\nu=M_\nu^TM_N^{-1}M_\nu$, we obtain \eq{olmenu} with
\be
a&=&x^2P+y^2Q,\quad b=\eta^3\(-x^2P+y^2Q\),\nn\\
c&=&yzQ,\quad f=z^2Q,\quad \Rw\nn\\
a+b\eta^2&=&2y^2Q.\labels{abcf21}\ee

\subsubsection{\underline{$\rho=3,\ C=$IR3+IR1}}
In this case,
{\small\be
G_C=\pmatrix{0&1&0\cr 1&0&0\cr 0&0&1\cr},\quad
 M_N^{-1}=\pmatrix{P&Q&R\cr Q&P&R\cr R&R&S\cr},\quad M_\nu:=\pmatrix{x&y&z\cr -y\eta^2&-x\eta^3&-z\cr w&-w\eta^3&0\cr}.\labels{case3}\\ \ee}
Note that the $3\x 3$ Dirac mass matrix $M_\nu$ here is a composite,
with its first two rows taken from the $M_\nu$ in \eq{case1}, and the last row taken from the 
first row (with a change of symbol) of the $M_\nu$ in \eq{case2}.

Computing from these $\ol M_\nu=M_\nu^TM_N^{-1}M_\nu$, we obtain \eq{olmenu} with
\be
a&=&(x^2+y^2\eta^4)P-2xy\eta^2Q+2(x-y\eta^2)wR+w^2S,\nn\\
b&=&2xyP-(x^2\eta+y^2)\eta^2Q+2(-x\eta^3+y)wR-w^2\eta^3S,\nn\\
c&=&z(x+y\eta^2)(P-Q),\quad f=2z^2(P-Q), \quad\Rw\nn\\
a+b\eta^2&=&(x+y\eta^2)^2(P-Q).\labels{abcf31}\ee
If we set $w=R=S=0$, we return to \eq{abcf3} as it should.

\subsubsection{\underline{$\rho=2,\ C=$IR3+IR2}}
In this case,
{\small\be
G_C=\pmatrix{0&1&0\cr 1&0&0\cr 0&0&-1\cr},\quad
M_N^{-1}=\pmatrix{P&Q&R\cr Q&P&-R\cr R&-R&S\cr},\quad M_\nu:=\pmatrix{x&y&z\cr -y\eta^2&-x\eta^3&-z\cr w&w\eta^3&u\cr}.
\labels{case4}\\ \ee}
The $3\x 3$ Dirac mass matrix $M_\nu$ here is also a composite,
with its first two rows taken from the $M_\nu$ in \eq{case1}, and the last row taken from the 
second row (with a change of symbol) of the $M_\nu$ in \eq{case2}.

Computing from these $\ol M_\nu=M_\nu^TM_N^{-1}M_\nu$, we obtain \eq{olmenu} with
\be
a&=&(x^2+y^2\eta^4)P-2xy\eta^2Q+2(x+y\eta^2)wR+w^2S,\nn\\
b&=&2xyP-(x^2\eta+y^2)\eta^2Q+2(x\eta^3+y)wR+w^2\eta^3S,\nn\\
c&=&z(x+y\eta^2)(P-Q)+\(2zw+u(x+y\eta^2)\)R+uwS,\nn\\
 f&=&2z^2(P-Q)+4zuR+u^2S,\quad\nn\\
a+b\eta^2&=&(x+y\eta^2)[(x+y\eta^2)(P-Q)+4wR+2w^2S.\labels{abcf32}\ee
If we set $u=w=R=S=0$, we return to \eq{abcf3} as it should.

\subsection{General Remarks}
In all the cases considered above, both in the charged-lepton and in the neutrino sectors, there are
more parameters than necessary to fix $\ol M_e$ and $\ol M_\nu$ in \eq{olmenu}, from which
all the low-energy leptonic physical quantities can be determined. This is specially so in case 4
above. As a result, the situation is very complicated and
there are many degenerate parameters that cannot be resolved by the
measured quantities. In the rest of this section, we will concentrate on studying the model $c=0$ in more detail,
where things become a bit simpler and more transparent.

\subsection{The Model $c=0$}
Consider $\ol M_N$ in \eq{olmenu} with $c=0$, a case already considered in Sec.~IIIA. 
Recall that all the experimental mixing angles can be obtained with this choice, together with 
$\delta_{CP}=0$.
We
would like to know what happens if we also impose a right-handed symmetry in the model.
For that purpose, let us concentrate on equations \eq{abcf3}, \eq{abcf21}, \eq{abcf31}, \eq{abcf32}, and \eq{modeldiag}.

It can be seen from these equations that $f=0$ often follows from $c=0$.  This
is undesirable because phenomenologically
 $m_2=|f|$ can never be zero, although either $m_1$ or $m_3$ may be.
In the first three cases, the only way to render $c=0$ and $f\not=0$ is to set
\be
x+y\eta^2=0\ {\rm in\ } \eq{abcf3}, \quad 
y=0\ {\rm in\ } \eq{abcf21}, \quad
x+y\eta^2=0\ {\rm in\ } \eq{abcf31}.
\labels{param}\ee
In all three cases, this automatically implies $m_3=|a+b\eta^2|=0$. This unexpected prediction
of $m_3=0$, and hence an inverted hierarchy, stems from the right-handed symmetries imposed on
these three cases, so at least in these cases, the imposition of right-handed symmetry does produce
additional predictions.

Beyond that, there are enough free
parameters left in each case to fit any $m_1$ and $m_2$. Specifically, for case 1 in \eq{abcf3},
\be
m_1=|a-b\eta^2|=|4x^2(P+Q)|,\quad m_2=|f|=|2z^2(P-Q)|.\labels{m121}\ee
For case 2 in \eq{abcf21},
\be
m_1=|a-b\eta^2|=|x^2P|,\quad m_2=|f|=|z^2Q|.\labels{m122}\ee
For case 3 in \eq{abcf31},
\be
m_1=|a-b\eta^2|=|4x^2(P+Q)+8xwR+2w^2S|,\quad m_2=|f|=|2z^2(P-Q)|.\labels{m123}\ee

The last case is more complicated because it involves by far the most number of parameters. 
There seems to be many ways to impose $c=0$ and $f\not=0$, and it seems quite possible to 
have $m_3\not=0$ with $c=0$. For now let 
us just concentrate on a simple
choice of parameters to render $c=0$, by letting $x+y\eta^2=0$ and $w=0$. With this
choice, $m_3=0$ as well,  and as before there are 
enough parameters left to fix 
\be
m_1=|a-b\eta^2|=2|a|=4|x^2(P+Q)|,\quad m_2=|f|=|2z^2(P-Q)+4zuR+u^2S|\labels{4m1m2}\ee
in many ways.

\section{Dynamical Models for $\Delta(150)$}
It was assumed in the last two sections that the family symmetry $\G=\Delta(150)$ broke down
to the residual symmetries generated by $F$ and $G$, leaving it unspecified how to achieve that dynamically.
In this section we make a further assumption that this is caused by the vacuum expectation values
of scalar fields coupled to  fermions in $\G$-invariant Yukawa interactions. 
We will assume these scalar fields $\f$,
called {valons}, to carry only $\G$-quantum numbers but no Standard Model quantum numbers. 
The Standard-Model quantum numbers are carried by the usual Higgs field $H$, so the compound
field $H\f$ appearing together in the Yukawa terms carries both Standard Model and family quantum
numbers. Since we are only interested in the family structure here, we will omit
all spacetime and Standard-Model details, and replace the Higgs field by its expectation value $\bk{v}$.
In the presence of the compound scalar field, the Yukawa terms have dimension 5, so the coupling constant is 
inversely proportional to some heavy scale $\Lambda$. The Yukawa terms can then be written in the form
\be \sum_{B,a,b,c}h_B\ \ol\psi^C_c\chi^A_a\f^B_b\ \bk{Cc|Bb,Aa} + {\rm h.c.},\labels{yukawa}\ee
in which the factor $\bk{v}/\Lambda$ has been absorbed into the Yukawa coupling constant $h_B$.
The fermion fields $\psi$ and $\chi$ are assumed to transform according to representations
$C$ and $A$, and the valon field $\f$  according to representation $B$. The indices $a,b,c$
are the components of $A,B,C$, and $\bk{Cc|Bb,Aa}$ are the Clebsch-Gordan
coefficients needed to render \eq{yukawa} $\G$-invariant.

Let $g_A, g_B, g_C$ be an element of $\G$ in representations $A, B, C$ respectively.
A basis is chosen so that these matrices are unitary, and real if the representation is real (see Appendix A).
That is also the basis in which the Clebsch-Gordan coefficients $\bk{Cc|Bb,Aa}$ are computed. 
From the covariance relation of Clebsch-Gordan coefficients \cite{Hamermesh}
\be \sum_{a',b',c'}(g_C^\dagger)_{cc'}(g_B)_{b'b}(g_A)_{a'a}\bk{Cc'|Bb',Aa'}=\bk{Cc|Bb,Aa},\labels{CG}\ee
 we see that the Yukawa term \eq{yukawa} is invariant under any $\G$-transformation
\be \psi\to g_C\psi,\ \chi\to g_A\chi, \phi\to g_B\phi. \labels{inv}\ee

To break the symmetry from $\G$ down to the residual symmetry, we impose vacuum expectation values
$\bk{\f^B}$, determined by a $\G$-invariant valon potential ${\cal V}(\phi)$ via the equation
of motion $\partial {\cal V}/\partial \f^{D^*}=0$, where $D^*$ is the complex conjugated
representation of $D$. In order to preserve the residual symmetry, the invariance \eq{inv} must remain
true for $g=F$ in the charged-lepton sector, and $g=G$  in the neutrino sector. This requires
\be F_B\bk{\f^B}_e=\bk{\f^B}_e,\quad G_B\bk{\f^B}_\nu=\bk{\f^B}_\nu\labels{align}\ee
for every $B$, where the subscripts $e$ and $\nu$ indicate in which sector the vacuum expectation
values apply. Incidentally, since the normalization of $\bk{\f^B}$ cannot be determined from \eq{align},
it is conventional to take out the energy unit with an arbitrary numerical constant 
and absorb them into the Yukawa constant $h_B$. What remains is usually called the vaccum {\it alignment}.

In other words, the alignment in the charged-lepton sector must be an invariant eigenvector of $F$,
and the alignment in the neutrino sector must be an invariant eigenvector of $G$. By invariant eigenvector,
we mean an eigenvector with eigenvalue $+1$.

The mass matrix after symmetry breaking can be read off from \eq{yukawa} to be
\be (M_e)_{ca}&=&\sum_{B,b}h_B\bk{\f^B_b}_e\bk{C'c|Bb,Aa}, \quad (M_\nu)_{ca}=\sum_{B,b}h_B\bk{\f^B_b}_\nu\bk{Cc|Bb,Aa},\nn\\
(M_N)_{ca}&=&\sum_{B,b}h_B\bk{\f^B_b}_\nu\bk{Cc|Bb,Ca},
\labels{mm}\ee
where the range of $B,b$ and the value of $h_B$ may be different for the three cases.
The $\G$-representation of the left-handed isodoublets is $A$, that of the right-handed charged leptons is $C'$,
and that of the right-handed neutrinos $N$ is $C$, with $C$ being a real representation.
It follows from \eq{CG} and \eq{align} that the constraint on the mass matrices so obtained,
\be M_e=F_{C'}^\dagger M_e F_A,\quad M_\nu=G_C^TM_\nu G_A, \quad M_N=G_C^TM_N G_C,
\labels{lrinv}\ee
is the same as \eq{diracmaj}, except that in \eq{diracmaj} $F_A$ was abbreviated as $F$ and
$G_A$ was abbreviated as $G$. In other words, dynamical models are consistent with both-handed
symmetries and may be considered as
a refinement of the latter.

Now that we have decided on the desired alignment to preserve residual symmetries, the question
is whether a potential ${\cal V}$, invariant under $\G$, can be devised to provide such alignments.
It turns out that every invariant potential does provide alignment solutions which are 
invariant eigenvectors of $F$, or $G$, or any other element $g$ of $\G$. The reason for this will be explained
below. 

Let us assume
${\cal V}={\cal V}^{(34)}+{\cal V}^{(2)}$ to be a polynomial consisting of two parts: ${\cal V}^{(34)}$
of degree 3 and 4, and ${\cal V}^{(2)}$ of degree 2. The quadratic part can be written as
\be {\cal V}^{(2)}=-\sum_B\mu_B^2\f^{B^*}\f^B,\labels{V2}\ee
where $\f^{B^*}=(\f^B)^*$ is $\f$ in the representation $B^*$.
 Accordingly, the equation of motion can be written as
\be \mu^2_D\f^D=\partial {\cal V}^{(34)}/\partial \f^{D^*}:={\cal Q}_D(\f^B).\labels{eom}\ee
The unspecified index $B$ in the argument of ${\cal Q}_D$ simply means that this may be a function
of $\f^B$ for several $B$'s. 
To obtain $\bk{\f^D}$ in the lowest order, we may simply consider the fields $\f^D$ in \eq{eom}
to be classical and identify them with $\bk{\f^D}$.

Let $g_B$ be an element of $\G$ in representation $B$. 
We will now show that if $\hat\f^B$ is a normalized invariant eigenvector of $g_B$, with $g_B\bk{\hat\f^B}=\bk{\hat\f^B}$, then $\f^D=\kappa\hat\f^D$ is a solution of \eq{eom} for every $D$ provided $\hat\f^D$
is the {\it unique} invariant eigenvector of $g_D$. The normalization factor $\kappa$ is a constant to be determined from \eq{eom} in a way to be explained.

To show this assertion, first note that
\be g_{D}{\cal Q}_D(\f^{B})={\cal Q}_D(g_{B}\f^{B})
={\cal Q}_D(\f^{B}),\ee
so ${\cal Q}_D$ is an invariant eigenvector of $g_D$, whatever ${\cal Q}_D$ and
the normalization of $\f^B$ are.  Suppose ${\cal Q}_D$ consists of a number of monomials $q^i_D$
of degree 2 and a number of monomials $c^j_D$ of degree 3, 
so that ${\cal Q}_D=\sum_i q^i_D+\sum_jc^j_D$.
Since every $q^i_D(\hat\phi^B)$ and every $c^j_D(\hat\f^B)$ is an invariant eigenvector of $g_B$, and since
this eigenvector is unique, we must have $q^i_D(\hat\f^B)=\a_i\hat\f^D$ and $c_j(\hat\f^B)=\b_j\hat\f^D$
for some constants $\a_i$ and $\b_j$ that are determined by the structure of $q^i$ and $c^j$. 
Since $q^i$ is quadratic and $c^j$ cubic, it follows that
 $q^i_D(\kappa\hat\f^B)=\a_i\kappa^2\hat\f^D$ and $c_j(\hat\f^B)=\b_j\kappa^3\hat\f^D$, hence
${\cal Q}_D(\kappa\hat\f^B)=(\sum_i \a_i\kappa^2+\sum_j\b_i\kappa^3)\hat\f^D$.
Consequently, if $\kappa\not=0$ is chosen to be the solution of the quadratic equation
\be \mu^2_D=\sum_i \a_i\kappa+\sum_j\b_i\kappa^2,\labels{eomk}\ee
then $\f^D=\kappa\hat\f^D$ is a solution of \eq{eom}. \ \rule[-.05cm]{.2cm}{.4cm}

Let us now denote the $\f$ that couples to fermions in the charged-lepton sector of 
\eq{yukawa} by $\f_e$, and the one that couples in the neutrino sector  by $\f_\nu$.
What  we want then is a solution with $\bk{\f^B_e}=\bk{\f^B}_e$ and $\bk{\f^B_\nu}=\bk{\f^B}_\nu$.
This can be accomplished if we choose the valon potential to be ${\cal V}={\cal V}_e(\f_e)+{\cal V}_\nu(\f_\nu)$,
 with ${\cal V}_e$ and ${\cal V}_\nu$ being any two $\G$-invariant potentials. However, if ${\cal V}$ includes
an interacting potential ${\cal V}_{e\nu}(\f_e,\f_\nu)$, then like spin-spin interaction, the resulting
alignments of $\bk{\f^B_e}$ and $\bk{\f^B_\nu}$ would be shifted. To be consistent with everything
up to now, we must assume either additional `shaping symmetry' can be found \cite{AF} to 
effectively forbid ${\cal V}_{e\nu}$,
or that Nature provides us with a small ${\cal V}_{e\nu}$ so that everything said up to now would be
approximately true. 

The dynamical model reviewed above is the most popular method to implement family symmetry,
but it requires the presence of valon fields which may or may not exist in Nature.
Since the Yukawa terms are of dimension 5, the theory is at best a lower-energy effective theory
used to explain leptonic masses and mixing. The heaviest leptonic mass is $m_\tau<2$ GeV, so one 
might expect the valon-degree of freedom to show up around there, but certainly there is no sign of
any of them up to the present. Moreover, if there are really new degrees of freedom present, one might
expect the Higgs coupling to fermion pairs to deviate from the Standard Model \cite{Lam3},
but hitherto there is no sign of that either in the 125-GeV Higgs candidate.

In contrast, either the left-handed or both-handed symmetry approach
to the mass matrix does not require the presence of valons, though they do leave open the question
of how the family symmetry is broken to the desired residual symmetry. In this sense they are similar
to the texture-zero approach to mass matrices, which for example yields the  celebrated
relation of Cabibbo angle in terms of ratio of quark masses \cite{FWZ},
but the origin and the location of the zeros generally remain somewhat obscured.

With the assumption of the existence of valons, the dynamical model does allow a better control
of symmetry, because it provides a physical explanation for the parameters appearing in $M_e, M_\nu$,
and $M_N$ as Yukawa couplings, which in principle can be directly measured if valons do exist.
Moreover, one might imagine the possibility of turning off certain Yukawa couplings on the grounds
that the corresponding valons do not exist, thereby obtaining relations between 
the parameters of mass
matrices, which may provide relations between physical quantities.

We do not expect this last possibility to materialize in the charged-lepton sector. 
As discussed in the last two sections,
there are three positive parameters present to determine the three charged-lepton masses.
If a relation between these three parameters exist, it would imply a relation between 
the charged-lepton masses.
The only known relation between these masses is the Koide relation \cite{Koide},
but that involves  square root of masses that cannot be obtained by this kind of 
Yukawa terms. For that reason we will concentrate on the neutrino sector in what follows.
The four cases considered below corresponds to the four cases studied in the last
section. In these cases, the left-handed neutrinos $\nu$ always transform in representation $A=$IR5, but 
the right-handed neutrinos $N$
transform according to different representation $C$ in different cases. With $\rho$ being the number of $N$'s,
the first two cases have $\rho=2$ and the last two have $\rho=3$.

Before going into the details of these four cases, it might be useful to provide a brief summary.
First of all, there are always enough Yukawa coupling constants $h_B$ in every case to provide the necessary
parameters $x, y, z, u, w, P, Q, R, S$ in the last section. Since some of these parameters are
redundant in determining the parameters $a, b, c, f$ of $\ol M_\nu$ needed to fix the measured quantities,
some of the Yukawa coupling constants can indeed be set to zero without giving rise to any
new physical predictions. Next, one might ask whether the model studied
in the last two sections with $c=0$ can be obtained by the absence of some valons, or equivalently, by setting
some Yukawa couplings to zero. The answer is no, though $c=0$ can be obtained by 
specific relations between two non-zero Yukawa parameters.

The Clebsch-Gordan coefficients needed to calculate the mass matrix in \eq{mm}
can be obtained with the help of GAP using \eq{CG}, which states that $\bk{Cc|Bc,Aa}$ is
the invariant eigenvector of $g_C^\dagger\otimes g_B^T\otimes g_A^T$. With that the computation of
the Clebsch-Gordan coefficients is reduced to an algebraic problem of finding the invariant
eigenvectors from the representations of group elements given by GAP.

If $\bra{Aa}, \bra{Bb}$ are the $a, b$ components of two vectors in the representation spaces $A$ and
 $B$ respectively, then
\be \bra{Cc}=\sum_{a,b}\bk{Cc|Bb,Aa}\bra{Bb}\bra{Aa}\labels{CG1}\ee
is the $c$ component of a vector in the representation space $C$. In what follows we will list a few
Clebsch-Gordan coefficients using \eq{CG1} and the following notation.
$a$ is a singlet in IR1, $b$ is a singlet in IR2, $(c,d )$ is a doublet in IR3, and $(e,f,g), (e',f',g'), (e'',f'',g'')$
are triplets in IR5, IR10, IR11 respectively. Subscripts may be used to distinguish two multiplets transforming
the same way. The product of a multiplet with $a$ always reproduces the multiplet, so we will skip
those relations. Some other useful ones are
{\small\be
\pmatrix{c\cr d\cr}&=&{1\over 2\rt}\pmatrix{(1-\rt)ee'+(1+\rt)ff'-2gg'\cr -(1+\rt)ee'+(-1+\rt)ff'+2gg'\cr},\nn\\ \nn\\
\pmatrix{c\cr d\cr}&=&{1\over 2\rt}\pmatrix{(-1-\rt)ee''+(-1+\rt)ff''+2gg''\cr (-1+\rt)ee''+(-1-\rt)ff''+2gg''\cr},\nn\\ \nn\\
\pmatrix{c\cr d\cr}&=&{1\over 2}\pmatrix{c_1c_2-c_1d_2-d_1c_2-d_1d_2\cr -c_1c_2-c_1d_2-d_1c_2+d_1d_2\cr},\nn\\ \nn\\
a&=&{1\over\rt}(ee''+ff''+gg''),\nn\\ 
b&=&{1\over\rt}(ee'+ff'+gg'),\nn\\ 
a&=&{1\over\rd}(c_1c_2+d_1d_2),\nn\\
b&=&{1\over\rd}(c_1d_2-c_2d_1).
\labels{CG2}\ee}

We will now proceed to study the four cases separately. Since there is really nothing new we can get that
way, all that we can do is to relate the parameters $x, y, z, u, w, P, Q, R, S$ of the last section to the
Yukawa coupling constants $h_B$, and to determine the relations between the $h_B$'s 
that can render $c=0$.

\subsection{\underline{$\rho=2,\ C=$IR3}}

\subsubsection{Dirac mass matrix $M_\nu$}
From Table 1, we know that IR5$\x$IR10 and IR5$\x$IR11 both contain IR3,  so $B$ in \eq{mm} consists 
of IR10 and IR11. The corresponding representation $G_B$ for the residual symmetry  is
{\small\be G_{10}=-G_{11}=\pmatrix{0&\eta^2&0\cr \eta^3&0&0\cr 0&0&1\cr}.\ee} 
The alignments obtained from \eq{align} 
are  $\bk{\f^{11}}=(-\eta^2,1,0)^T$ in one case, and 
an arbitrary linear combination of $\bk{\f^{10a}}=(\eta^2,1,0)^T$ and $\bk{\f^{10b}}=(0,0,1)^T$
in the other case. The reason why IR11 has one invariant eigenvector and IR10 has 2 is because 
 $\det(G_{11})=+1$ but $\det(G_{10})=-1$. There are  two Yukawa coupling constants
$h_{10}$ and $h_{11}$, but since the combination of the two invariant eigenvectors of IR10 is 
arbitrary, there are effectively three unknown coefficients, 
which we will refer to as $h_{10a}, h_{10b}$, and $h_{11}$.  Using \eq{mm} and \eq{CG2}, we find
that $M_\nu$ is of the form given in \eq{case1}, with
\be
x={1\over 2\sqrt{3}}\(h_{10a}x^{10a}+h_{11}x^{11}\), \quad y={1\over 2\sqrt{3}}\(h_{10a}y^{10a}+h_{11}y^{11}\), \quad z=-{1\over\sqrt{3}}h_{10b},\labels{xyzI}\ee
where
\be
x^{10a}&=&  \eta^2(1-\rt), \quad y^{10a}=1+\rt,\nn\\
x^{11}&=& \eta^2(1+\rt), \quad
          y^{11}=-1+\rt.\labels{xy1011}\ee
Using them to compute $(x+y\eta^2)$, which is proportional to $c$, we get
\be (x+y\eta^2)^{10a}&=&2\eta^2,\quad (x+y\eta^2)^{10b}=0,\nn\\
(x+y\eta^2)^{11}&=&2\eta^2\rt,\ee
hence
\be \sqrt{3}(x+y\eta^2)^{10a}-(x+y\eta^2)^{11}=0.\labels{yy}\ee
Thus if $h_{11}=-\sqrt{3}h_{10a}$, then  $x+y\eta$ is zero, 
giving $c=0$ and the model considered in the last two sections. As discussed in the
last section, $m_3=0$ follows automatically in this case.

\subsubsection{Majorana Mass Matrix $M_N$}
There is no difficulty in computing the Majorana mass matrix $M_N$ as we shall see, 
but $M_N^{-1}$ is needed for $\ol M_\nu$, so 
a matrix inversion has to be performed. Since $M_N$ and $M_N^{-1}$
satisfy the same symmetry constraint, \eq{diracmaj} and \eq{olmenu2},  their solutions
must have the same form.
Henceforth we will use the form of $M_N^{-1}$
and the same letters {\it with a prime} to parametrize $M_N$. 
In the case of \eq{case1}, the relation between these two sets of parameters are
\be P=P'/D_1,\quad Q=-Q'/D_1,\quad D_1:={P'}^{2}-{Q'}^2.\labels{inverse1}\ee

From Table 1, we see that IR3$\x$IR$B$=IR3 if $B$=1, 2, or 3. The residual symmetry $G_B$
for $B=3$ is given in \eq{G3}, and the other two, $G_1=-G_2=+1$, are given in Table 1. The
alignments are  $\bk{\f^3}=(1,1)^T, \bk{\f^1}=1$, and $\bk{\f^2}=0$, so only $B=$IR1 and IR3 
contribute. The 
Majorana mass matrix $M_N^{-1}$ of the form given in \eq{case1}, with 
$P'=h_1$ and $Q'=-h_3$. The parameters $P, Q$ for $M_N^{-1}$ can be obtained from \eq{inverse1}.

\subsection{\underline{$\rho=2,\ C=$IR1+IR2}}

\subsubsection{Dirac mass matrix $M_\nu$}
From Table 1, we know that IR5$\x$IR11 contains IR1,  and IR5$\x$IR10
contains IR2,  so once again $B$ in \eq{mm} consists 
of IR10 and IR11. The corresponding representation $G_B$ and the alignments are the same as
those given in Sec.~VA.  The Dirac mass matrix $M_\nu$ is the sum of $M_\nu^a$ and $M_\nu^b$,
obtained using respectively alignments $\bk{\f^{10a}}$ and $\bk{\f^{10b}}$ for IR2, and alignment
$\bk{\f^{11}}$ for IR1. $M_\nu$ can be calculated from \eq{mm} and \eq{CG2}. It is of the form in
\eq{case2}, with
\be
x&=&-h_{11}\eta^2/\rt,\quad y=h_{10a}\eta^2/\rt, \quad z=h_{10b}/\rt.\labels{xyzII}\ee
From \eq{abcf21}, we see that in order to reproduce the model with $c=0$ and $f\not=0$, we need
to have $y=0$ and hence $h_{10a}=0$.

\subsubsection{Majorana Mass Matrix $M_N$}
Since IR1$\x$IR1=IR1 and IR2$\x$IR1=IR2, if we let the two Yukawa coupling constants be
$h_1$ and $h_1'$ respectively, then the Majorana Matrix $M_N^{-1}$ is given by \eq{case2},
with $P=h_1^{-1}$ and $Q=h_1'^{-1}$.

\subsection{\underline{$\rho=3,\ C=$IR3+IR1}}

\subsubsection{Dirac mass matrix $M_\nu$}
As remarked in Sec.~IVB3, the $M_\nu$ here is a composite, with the first two rows given bt the $M_\nu$
in VA, and the last row given by the first row of the $M_\nu$ in VB. 
The parameters $x, y, z, w$ in \eq{case3} can now be copied from \eq{xyzI} and \eq{xyzII} to be
\be x&=&{1\over 2\sqrt{3}}\(h_{10a}x^{10a}+h_{11}x^{11}\), \quad y={1\over 2\sqrt{3}}\(h_{10a}y^{10a}+h_{11}y^{11}\),\nn\\ z&=&-{1\over\sqrt{3}}h_{10b}, \quad
w=-{1\over\rt}h'_{11}\eta^2,\labels{xyzwIII}\ee
where $x^{10a}, y^{10a}, x^{11}, y^{11}$ are given in \eq{xy1011}, $h_B$ is the Yukawa coupling
to IR3, and $h'_B$ is the Yukawa coupling to IR1.

Again, as in VA, because of \eq{yy}, to render $c=0$ we need to have $h_{11}=-\sqrt{3}h_{10a}$.

\subsubsection{Majorana Mass Matrix $M_N$}
Both $M_N$ and $M_N^{-1}$ are of the form \eq{case3}, with their parameters related by
\be
P&=&(P'S'-{R'}^2)/D_{3a},\quad Q=-(Q'S'-{R'}^2)/D_{3a},\quad D_{3a}:=(P'-Q')D_{3b},\nn\\
R&=&-R'/D_{3b},\quad S=(P'+Q')/D_{3b},\quad D_{3b}:=(P'+Q')S'-2{R'}^2.
\labels{inverse3}\ee
The parameters for $M_N$ can be partially copied from VA and VB:
\be
P'&=&h_1,\quad Q'=-h_3, \quad R'=h_3'/\rd, \quad S'=h'_1, \labels{N3}\ee
where the couplings are respectively for IR3$\x$IR1$\to$IR3,  IR3$\x$IR3$\to$IR3,
IR3$\x$IR3$\to$IR1, and IR1$\x$IR1$\to$IR1.

\subsection{\underline{$\rho=3,\ C=$IR3+IR2}}

\subsubsection{Dirac mass matrix $M_\nu$}
As remarked in Sec.~IVB4, the $M_\nu$ here is a composite, with the first two rows given bt the $M_\nu$
in VA, and the last row given by the second row of the $M_\nu$ in VB. 
The parameters $x, y, z, w$ in \eq{case3} can now be copied from \eq{xyzI} and \eq{xyzII} to be
\be x&=&{1\over 2\sqrt{3}}\(h_{10a}x^{10a}+h_{11}x^{11}\), \quad y={1\over 2\sqrt{3}}\(h_{10a}y^{10a}+h_{11}y^{11}\),\nn\\ z&=&-{1\over\sqrt{3}}h_{10b}, \quad
w=-{1\over\rt}h'_{11}\eta^2,\labels{xyzwIV}\ee
where $x^{10a}, y^{10a}, x^{11}, y^{11}$ are given in \eq{xy1011}, $h_B$ is the Yukawa coupling
to IR3, and $h'_B$ is the Yukawa coupling to IR1.

Again, as in VA, because of \eq{yy}, to render $c=0$ we need to have $h_{11}=-\sqrt{3}h_{10a}$.

\subsubsection{Majorana Mass Matrix $M_N$}
Both $M_N$ and $M_N^{-1}$ are of the form \eq{case4}, with their parameters related by
\be
P&=&(P'S'-{R'}^2)/D_{4a},\quad Q=-(Q'S'+{R'}^2)/D_{4a},\quad D_{4a}:=(P'+Q')D_{4b},\nn\\
R&=&-R'/D_{4b},\quad S=(P'-Q')/D_{4b},\quad D_{4b}:=(P'-Q')S'-2{R'}^2.
\labels{inverse4}\ee
The parameters for $M_N$ can be partially copied from VA and VB:
\be
P'&=&h_1,\quad Q'=-h_3, \quad R'=h_3'/\rd, \quad S'=h'_1, \labels{NN3}\ee
where the couplings are respectively for IR3$\x$IR1$\to$IR3,  IR3$\x$IR3$\to$IR3,
IR3$\x$IR3$\to$IR2, and IR2$\x$IR1$\to$IR2.

\section{Neutrino Mixing of $\Delta(600)$}

A parallel was mentioned in the Introduction between  symmetry considerations
of the old and the new mixing data. The zero reactor
angle and the maximal atmospheric angle of the old data can be explained by a $S_3$ symmetry,
a group  generated by the matrices $\tilde F=\diag(1,\o,\o^2)$ and an order-2 unitary 
matrix $\tilde G$ 
with an invariant eigenvector $(0,1,-1)^T$.
In order to have a chance to explain the solar angle by symmetry as well, we need  another order-2
unitary matrix $\tilde G'$ which commutes with $\tilde G$. Such a matrix
with an invariant eigenvector $(1,1,1)^T$ does exist and is contained in the group $S_4\supset S_3$. These two 
invariant eigenvectors together give rise to tribimaximal mixing and the correct solar angle.

With the new data of a non-zero reactor angle and possibly non-maximal atmospheric mixing, we saw in
the last three sections that they can now be explained by the symmetry group $\Delta(150)$, 
a group generated
by $F$ and  $G$ of \eq{FGv} and \eq{FGf}. Again, to have a chance to explain the solar angle by symmetry, we need to find another order-2
unitary operator $G'$ which commutes with $G$. The smallest group containing 
such an operator as well as the subgroup $\Delta(150)$ is $\Delta(600)$.

This motivation for the full-symmetry group $\Delta(600)$ is based completely on symmetry considerations.
There is another motivation based on dynamics which also has a parallel with the old data. 
The most popular dynamical
model for the old data is based on the symmetry group $A_4$ \cite{A4}, generated by $\tilde F$
and $\tilde G'$ above. If its neutrino coupling to the ${\bf 1'}$ and ${\bf 1''}$ valons are dynamically
suppressed, then the solar angle comes out correct, the matrix $\tilde G$ becomes also a symmetry,
and the group $A_4$ is promoted to $S_4$.

The analogy of $A_4$ with the new data is $\Delta(150)$. By setting the parameter $c$ dynamically to zero,
we saw in the last three sections that the solar angle can also be explained.
In the GAP basis where the generators of $\Delta(150)$ are the $F, G$ of \eq{FGv} and \eq{FGf}, the mixing
vector for the model with $c=0$  is $(0,0,1)^T$. It turns out this is just the invariant
eigenvector of the operator $G'$ in the group $\Delta(600)$.

To see that,  let us first find out what $G'$ is. It must be an order-2 element of $\Delta(600)$ that
commutes with $G$ of \eq{FGf}. In terms of the generators of $\Delta(600)$ given in Sec.~IIB, 
$G=f_1c_3^2c_4^4$. According to \eq{f600d}, $f_1, c_3, c_4$ all commute with the order-2 element $d_3$,
so a natural candidate for $G'$ is a three-dimensional representation of $d_3$. From Table 5,
there are 18 three-dimensional representations. We must choose one for $f$ and $g$ which reproduces
$F$ and $G$ in \eq{FGv}. This turns out to be IR14. The representation of $d_3$ in IR14
is $G'=\diag(-1,-1,1)$, which does possess the desired invariant eigenvector $(0,0,1)^T$.

We know from Sec.~IIB that $F$ and $G'$ generate a $A_4$ subgroup of $\Delta(600)$, and we already
know that $F$ and $G$ generate the subgroup $\Delta(150)$. In this way $\Delta(600)$ for the new
data is like $S_4$ for  the old data, which contains and are generated by the matrices $\tilde F, \tilde G$,
and ${\tilde G}'$. In fact, if we use $V$ of \eq{V} to convert the $F$-diagonal to the GAP representation, 
$\tilde F$ is the same as $F$, $\tilde G'$ is the same as $G'$, though $\tilde G$ and $G$ are different.
$\tilde F, \tilde G$ generate an $S_3$ subgroup of $S_4$, and $F, G$ generate an $\Delta(150)$
subgroup of $\Delta(600)$.

\subsection{Left-Handed Mass Matrices for $\Delta(600)$}
Apply the symmetry constraints 
\be
F^\dagger \ol M_e F=\ol M_e,\quad G^T\ol M_\nu G=\ol M_\nu, \quad {G'}^T\ol M_\nu G'=\ol M_\nu
\labels{sym600}\ee
to the Hermitian $\ol M_e$ and the symmetric $\ol M_\nu$, we get the same solution $\ol M_e$
as shown in \eq{olmenu}. For $\ol M_\nu$, the extra symmetry constraint provided by $G'$ reduces
the form in \eq{olmenu} to the form in \eq{olmnu0} for the special model. Hence the $c=0$ model
of $\Delta(150)$ can be realized by the increased symmetry of $\Delta(600)$,  just like the zero
${\bf 1'}, {\bf 1''}$ model of $A_4$ can be realized by the increase symmetry of $S_4$.

\section{Summary}
The symmetry group $\Delta(150)$ is applied to neutrino mixing to obtain the correct reactor and atmospheric
angles. Three methods are used to implement this symmetry: by imposing it to the left-handed fermions alone,
by imposing it to both the left-handed and the right-handed fermions, and by constructing dynamical models
based on the symmetry. The relative merits of these three methods are discussed. 
Generally speaking, as far as mixing parameters are concerned, it is sufficient to use
the left-handed symmetry. A simple
model with the parameter $c=0$ has been discussed in some detail. 
This model gives the correct solar angle and can be reproduced from the 
enlarged group $\Delta(600)$ by symmetry alone.

\newpage
 \appendix
\section{Real Representations of $\Delta(150)$}
We see from Table 1 in Sec.~II that the characters of IR1, 2, 3, 12, 13 are real. As noted in
Sec.~IIA, these are actually real rather than quaternionic representations because 
their Frobenius-Schur indices are  $+1$.
Let $f_i(n)$ denote the representation of the generator $f_i$ in IR$n$. For IR1 and IR2, the irreducible representation is the same as the character, which
is explicitly real. For IR3, the representation given in GAP 
{\small
\be
f_1(3)&=&\pmatrix{ 0& 1\cr  1& 0\cr},\quad f_2(3)=\pmatrix{ \o&0\cr 0&\o^2\cr},\nn\\ \nn\\ 
 f_3(3)&=&\pmatrix{ 1&0\cr 0&1\cr},\quad f_4(3)=\pmatrix{1&0\cr 0&1\cr},\labels{fa3}\ee}
is not all real, and neither is the case for IR12 and IR13:
{\footnotesize
\be f_1(12)&=&\pmatrix{0& 0& 0& 0& 0& 1 \cr   0& 0& 0& 0& 1& 0 \cr   0& 0& 0& 1& 0& 0 \cr 
        0& 0& 1& 0& 0& 0 \cr   0& 1& 0& 0& 0& 0 \cr   1& 0& 0& 0& 0& 0 \cr},\quad
f_2(12)=\pmatrix{ 0& 0& 1& 0& 0& 0 \cr   1& 0& 0& 0& 0& 0 \cr   0& 1& 0& 0& 0& 0 \cr 
        0& 0& 0& 0& 0& 1 \cr   0& 0& 0& 1& 0& 0 \cr   0& 0& 0& 0& 1& 0 \cr},\nn\\ \nn\\
f_3(12)&=&\pmatrix{ 1& 0& 0& 0& 0& 0 \cr   0& \eta ^3& 0& 0& 0& 0 \cr 
        0& 0& \eta ^2& 0& 0& 0 \cr   0& 0& 0& \eta ^2& 0& 0 \cr 
        0& 0& 0& 0& \eta ^3& 0 \cr   0& 0& 0& 0& 0& 1 \cr},\quad
f_4(12)=\pmatrix{ \eta & 0& 0& 0& 0& 0 \cr   0& \eta ^3& 0& 0& 0& 0 \cr 
        0& 0& \eta & 0& 0& 0 \cr   0& 0& 0& \eta ^2& 0& 0 \cr 
        0& 0& 0& 0& \eta ^4& 0 \cr   0& 0& 0& 0& 0& \eta ^4 \cr};
\labels{f12}\ee}

{\footnotesize
\be f_1(13)&=&\pmatrix{   0& 0& 0& 0& 0& 1 \cr   0& 0& 0& 0& 1& 0 \cr   0& 0& 0& 1& 0& 0 \cr 
        0& 0& 1& 0& 0& 0 \cr   0& 1& 0& 0& 0& 0 \cr   1& 0& 0& 0& 0& 0 \cr},\quad
f_2(13)=\pmatrix{ 0& 0& 1& 0& 0& 0 \cr   1& 0& 0& 0& 0& 0 \cr   0& 1& 0& 0& 0& 0 \cr 
        0& 0& 0& 0& 0& 1 \cr   0& 0& 0& 1& 0& 0 \cr   0& 0& 0& 0& 1& 0 \cr},\nn\\ \nn\\
f_3(13)&=&\pmatrix{  \eta & 0& 0& 0& 0& 0 \cr   0& \eta ^4& 0& 0& 0& 0 \cr   0& 0& 1& 0& 0& 0 \cr 
        0& 0& 0& 1& 0& 0 \cr   0& 0& 0& 0& \eta ^4& 0 \cr 
        0& 0& 0& 0& 0& \eta  \cr},\quad
f_4(13)=\pmatrix{ \eta & 0& 0& 0& 0& 0 \cr   0& \eta ^2& 0& 0& 0& 0 \cr 
        0& 0& \eta ^2& 0& 0& 0 \cr   0& 0& 0& \eta ^3& 0& 0 \cr 
        0& 0& 0& 0& \eta ^4& 0 \cr   0& 0& 0& 0& 0& \eta ^3 \cr }.
\labels{f13}\ee}

Although these matrices may not be real, they are all unitary.
In that case \cite{Hamermesh},  there is a unitary and symmetric matrix $B(n)$ which  renders $B(n)f_i(n)B(n)^{-1}=\ol f_i(n)$  explicitly real. Moreover,  $S(n)f_i(n)S(n)^{-1}=f_i(n)^*$
when $S(n)=B(n)^2$. Our task is to find $S(n)$ and $B(n)$, and here is how.

To simplify writing, let us drop the dependence on $n$ and $i$. If $v$ is an eigenvector
of $\ol f$ with eigenvalue $\l$, then $v^*$ is an eigenvector of $\ol f$ with eigenvalue
$\l^*$. From $BfB^{-1}=\ol f=\ol f^{\ *}$, we conclude that $u:=B^{-1}v$ is an eigenvector of $f$ with eigenvalue $\l$,
and $w:=B^{-1}v^*$ is an eigenvector of $f$ with eigenvalue $\l^*$. If $\l$ is complex and non-degenerate,
then $v=Bu=(Bw)^*$, hence 
\be Su=w^*\labels{suw}\ee 
because $B$ is unitary and symmetric. Even if $\l$ is degenerate or real, $(Su)^*$ is still an
eigenvector of $f$ with eigenvalue $\l$. Together with \eq{suw}, these relations can be
used to obtain $S(n)$ from $u_i(n)$ and $w_i(n)$.

Once $S$ is known, $B=\sqrt{S}$ can be computed as follows. Suppose $e_a$  is the normalized
eigenvector of $S$
with eigenvalue $\s_a$, then 
\be S=\sum_a\s_ae_ae_a^\dagger,\quad \Rw\ B=\sum_a\sqrt{\s_a}e_ae_a^\dagger.\labels{sb}\ee
Since each $\sqrt{\s_a}$ has two values, there can be many $B$'s, but any one of them would do.

\subsection{$n=3$}
Only $f_2(3)$ in \eq{fa3} is complex. $u=(1,0)^T$ is its eigenvector with eigenvalue $\l=\o$, and $w=(0,1)^T$
is its eigenvector with eigenvalue
$\l^*=\o^2$. Thus $Su=w^*$ implies 
{\small\be
S(3)=\pmatrix{0&1\cr 1&0}=e_1e_1^\dagger-e_2e_2^\dagger,\labels{s3}\ee}
with $e_1=(1,1)^T/\rd$ and $e_2=(1,-1)^T/\rd$.
Hence
{\small\be
B(3)=e_1e_1^\dagger+ie_2e_2^\dagger=\pmatrix{x&x^*\cr x^*&x\cr},\labels{b3}\ee}
where $x=(1+i)/2=\exp(\pi i/4)/\rd$.
The real representations $B(3)f_i(3)B(3)^{-1}=\ol f_i(3)$ are given by
{\small\be
\ol f_1(3)&=&\pmatrix{0&1\cr 1&0\cr},\quad \ol f_2(3)=-{1\over 2}\pmatrix{1&2c_{7/12}\cr -2c_{7/12}&1\cr},
\nn\\ \nn\\
\ol f_3(3)&=&\ol f_4(3)=\pmatrix{1&0\cr 0&1\cr},\labels{bf3}\ee}
where
\be
c_r:=\cos(2\pi i r),\quad s_r:=\sin(2\pi ir).\labels{csr}\ee

We computed $S(3)$ using \eq{suw} to illustrate the general technique. In the present case, it
can be obtained without this heavy machinery. Since $f_2(3)^*$ is obtained
from $f_2(3)$ by interchanging the (11) and the (22) entries, it is obvious that $S(3)$ must be
given by \eq{s3}.

\subsection{$n=12$}
%The complex eigenvalues of $f_i(12)$ in \eq{f12} come from $i=3$ and 4, with their eigenvalues $\l$ and
%eigenvectors $u$ listed in Table A1.
%
%$$\ba{|c|c|c|c||c|c|c|c||c|c|c|c|}\hline
%\#&i&\l&u^T&\#&i&\l&u^T&\#&i&\l&u^T\\ \hline
%A1&3&1& [ 1, 0, 0, 0, 0, 0 ]& C1&3&\eta^3&[ 0, 1, 0, 0, 0, 0 ]  &E&4&\eta^2&[ 0, 0, 0, 1, 0, 0 ]\\
%A2&&&[ 0, 0, 0, 0, 0, 1 ]&C2&&&[ 0, 0, 0, 0, 1, 0 ]&F&4&\eta^3&[ 0, 1, 0, 0, 0, 0 ]\\
%B1&3&\eta^2& [ 0, 0, 1, 0, 0, 0 ] &D1&4&\eta&[ 1, 0, 0, 0, 0, 0 ] &G1&4&\eta^4&[ 0, 0, 0, 0, 1, 0 ]\\ 
%B2&&&[ 0, 0, 0, 1, 0, 0 ]&D2&&&[ 0, 0, 1, 0, 0, 0 ]&G2&&&[ 0, 0, 0, 0, 0, 1 ]\\
%\hline\ea$$
%\vskip.5cm
%\bc Table A1. Eigenvalues $\l$ and eigenvectors $u$ of $f_i(12)$ for $i=3, 4$\ec

%We see from Tabel A1 that $S$ maps (the $u$ in) E to F (which is the same as B2 to C1), hence
%it also maps B1 to C2 (which is the same as D2 to G1), and therefore also D1 to G2. Thus

Since the complex matrices $f_3(12)$ and $f_4(12)$ are diagonal, it is again easy to obtain $S(12)$
directly as follows. $f_3(12)^*$ is obtained from $f_3(12)$
by interchanging the (11) entry either with the (22) or the (55) entry, but comparing $f_4(12)^*$
with $f_4(12)$ tells us that it must be the (55) and not the (22) entry. From $f_3$ we then know that
(22) must swap with (66) to get $f_3^*$, then from $f_4$ we know that (33) must swap with (44),
thereby obtaining

{\footnotesize\be S(12)=\pmatrix{0 &0 &0 &0 &0 &1 \cr 0 &0 &0 &1 &0 &0 \cr 0 &0 &0 &0 &1 &0 \cr 0 &1 &0 &0 &0 &0 \cr 0 &0 &1 &0 &0 &0\cr 1 &0 &0 &0 &0 &0\cr}.
\labels{s12}\ee}

Taking the square root of $S(12)$, we get
{\footnotesize\be B(12)=\pmatrix{ x& 0& 0& 0& 0& x^* \cr 
   0& x& 0& x^*& 0& 0 \cr  
   0& 0& x& 0& x^*& 0 \cr  
   0& x^*& 0& x& 0& 0 \cr  
   0& 0& x^*& 0& x& 0 \cr  
  x^*& 0& 0& 0& 0& x }.
\labels{b12}\ee}
The real representations $B(12)f_i(12)B(12)^{-1}=\ol f_i(12)$ are then given by
{\footnotesize\be
\ol f_1(12)&=&\pmatrix{  0 &  0 &  0 &  0 &  0 &  1  \cr   0 &  0 &  0 &  0 &  1 &  0  \cr   0 &  0 &  0 &  1 &  0 &  0  \cr  
       0 &  0 &  1 &  0 &  0 &  0  \cr  0 &  1 &  0 &  0 &  0 &  0  \cr   1 &  0 &  0 &  0 &  0 &  0 \cr},  \quad
\ol f_2(12)=\pmatrix{  0 &  0 &  1 &  0 &  0 &  0  \cr   1 &  0 &  0 &  0 &  0 &  0  \cr   0 &  1 &  0 &  0 &  0 &  0  \cr  
       0 &  0 &  0 &  0 &  0 &  1  \cr   0 &  0 &  0 &  1 &  0 &  0  \cr  0 &  0 &  0 &  0 &  1 &  0 \cr }, \nn\\ \nn\\
\ol f_3(12)&=&\pmatrix{   1 &  0 &  0 &  0 &  0 &  0  \cr  
       0 &  c_{2/5} &  0 & c_{3/20} &  0 &  0  \cr  
       0 &  0 &  c_{2/5} &  0 &  c_{7/20} &  0  \cr  
      0 &  c_{7/20} &  0 &  c_{2/5} &  0 &  0  \cr  
       0 &  0 &  c_{3/20} &  0 &  c_{2/5} &  0  \cr  
       0 &  0 &  0 &  0 &  0 &  1 \cr }, \quad
\ol f_4(12)=\pmatrix{  c_{1/5} &  0 &  0 &  0 &  0 &  c_{9/20}  \cr  
       0 &  c_{2/5} &  0 &  c_{3/20} &  0 &  0  \cr  
       0 &  0 &  c_{1/5} &  0 &  c_{9/20} &  0  \cr  
       0 & c_{7/20} &  0 &  c_{2/5} &  0 &  0  \cr  
      0 &  0 &  c_{1/20} &  0 &  c_{1/5} &  0  \cr  
       c_{1/20} &  0 &  0 &  0 &  0 &  c_{1/5} \cr }.
\labels{barf12}\ee}

\subsubsection{$n=13$}

The computation is similar to the $n=12$ case. The results are

{\footnotesize\be
S(13)&=&\pmatrix{0&0&0&0&1&0\cr 0&0&0&0&0&1\cr 0&0&0&1&0&0\cr 0&0&1&0&0&0\cr 1&0&0&0&0&0\cr 0&1&0&0&0&0\cr},\quad 
B(13)=\pmatrix{ x & 0& 0& 0& x^*& 0 \cr 
   0& x& 0& 0& 0& x^* \cr 
   0& 0& x & x^*& 0& 0 \cr 
   0& 0& x^*& x & 0& 0 \cr 
   x^*& 0& 0& 0& x & 0 \cr 
   0& x^*& 0& 0& 0& x  \cr},\nn\\ \nn\\
\ol f_1(13)&=&\pmatrix{  0& 0& 0& 0& 0& 1 \cr  0& 0& 0& 0& 1& 0 \cr  0& 0& 0& 1& 0& 0 \cr 
       0& 0& 1& 0& 0& 0 \cr  0& 1& 0& 0& 0& 0 \cr  1& 0& 0& 0& 0& 0 \cr},\quad
\ol f_2(13)=\pmatrix{   0& 0& 1& 0& 0& 0 \cr  1& 0& 0& 0& 0& 0 \cr  0& 1& 0& 0& 0& 0 \cr 
       0& 0& 0& 0& 0& 1 \cr  0& 0& 0& 1& 0& 0 \cr  0& 0& 0& 0& 1& 0 \cr},\nn\\ \nn\\
\ol f_3(13)&=&\pmatrix{  c_{1/5}& 0& 0& 0& c_{9/20}& 0 \cr 
       0& c_{1/5}& 0& 0& 0& -c_{9/20} \cr 
       0& 0& 1& 0& 0& 0 \cr  0& 0& 0& 1& 0& 0 \cr 
       -c_{9/20}& 0& 0& 0& c_{1/5}& 0 \cr 
       0& c_{9/20}& 0& 0& 0& c_{1/5} \cr},\quad
\ol f_4(13)=\pmatrix{  c_{1/5}& 0& 0& 0& c_{9/20}& 0 \cr 
       0& c_{2/5}& 0& 0& 0& c_{7/20} \cr 
       0& 0& c_{2/5}& c_{7/20}& 0& 0 \cr 
       0& 0& c_{3/20}& c_{2/5}& 0& 0 \cr 
       -c_{9/20}& 0& 0& 0& c_{1/5}& 0 \cr 
       0& c_{3/20}& 0& 0& 0& c_{2/5} \cr}.
\labels{sbbarf13}\ee}

%%%%%%%%%%%%%%%%%%%%%
%%%%%%%%%%%%%%%%%%%%%
%%%%%%%%%%%%%%%%%%%%%
%%%%%%%%%%%%%%%%%%%%%

\edoc
%%%%%%%%%%%%%%%%%%%%%%%